# Nanopore-Based DNA Sequencing Sensors and CMOS Readout Approaches

Mehdi Habibi

*Department of Electrical Engineering, Sensors and Interfaces Research Group, University of Isfahan, Isfahan, Iran*

Yunus Dawji, Ebrahim Ghafar-Zadeh, Sebastian Magierowski

*Department of Electrical Engineering and Computer Science, York University, Toronto, Canada*

**Abstract**

Purpose

Nanopore-based molecular sensing and measurement, specifically Deoxyribonucleic acid (DNA) sequencing, is advancing at a fast pace. Some embodiments have matured from coarse particle counters to enabling full human genome assembly. This evolution has been powered not only by improvements in the sensors themselves, but also in the assisting microelectronic Complementary Metal Oxide Semiconductor (CMOS) readout circuitry closely interfaced to them. In this light, this paper reviews established and emerging nanopore-based sensing modalities considered for DNA sequencing and CMOS microelectronic methods currently being used.

Design/methodology/approach

Readout and amplifier circuits which are potentially appropriate for conditioning and conversion of nanopore signals for downstream processing are studied. Furthermore, arrayed CMOS readout implementations are focused on and the relevant status of the nanopore sensor technology is reviewed as well.

Findings

Ion channel nanopore devices have properties unique compared with other electrochemical cells. Currently biological nanopores are the only variants reported which can be used for actual DNA sequencing. The translocation rate of DNA through such pores, the current range at which these cells operate on and the cell capacitance effect, all impose the necessity of using low noise circuits in the process of signal detection. The requirement of using in-pixel low noise circuits in turn tends to impose challenges in the implementation of large size arrays.

Originality/value

The study presents an overview on the readout circuits used for signal acquisition in electrochemical cell arrays and investigates the specific requirements necessary for implementation of nanopore type electrochemical cell amplifiers and their associated readout electronics.

## 1 Introduction

Nanopores are nanometer-sized apertures that, by virtue of their minute physical dimensions, confine individual molecules to the very fine spatial locale defined by their opening. This localization is a critical attribute of these sensors and is perhaps their most defining feature: single-molecule-sensing capability. Generally speaking, the operation of nanopores involves the establishment of some known external excitation to the pore such as a direct current (DC) ionic flowing through its passage; this is followed by the introduction of a 'test' molecule that interacts with the sensor (e.g. passes through it) over time and thus perturbs the pore's pre-established excitation in proportion to the interacting molecule's make up. The perturbed time-series is measured and subsequently used to identify the structure of the test molecule.

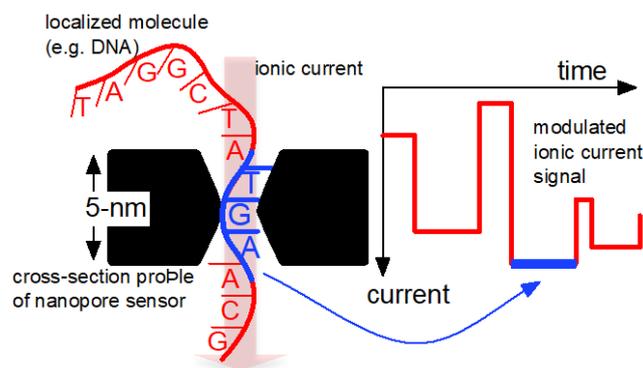

Figure 1: An abstract depiction of a DNA molecule localized within a nanopore (the profile of its cross-section) in the process of moving through the sensor while perturbing an ionic current. The result is a modulated time-series signal as drawn in simplified form. The highlighted blue level is indicative of a DNA segment translocating the sensor at the corresponding time.

Arguably, the most prominent success to-date of nanopore-based molecular measurement resides in its use within commercial DNA strand sequencing machines (Brown & Clarke 2016). An abstract illustration of such a nanopore-based process is pictured in Figure 1. As indicated, such sequencing systems thread (translocate) DNA through the nanopore; in the process a background ionic current is modulated in some – usually complicated – relationship to the string of *base* molecules (represented by the 'A, C, G, T' text) comprising the DNA. Ensuing signal processors convert the modulated signal to the DNA's base-equivalent text.

Since the first published proposals and demonstrations of the nanopore approach to molecular sensing in the mid-90's (Church et al. 1998, Kasianowicz et al. 1996) using α-hemolysin (αHL) proteins as the sensing aperture, the literature has seen a vast proliferation in structures, methods, and applications of this technique. Among these, excellent summaries of the state-of-the-art have been presented as well (Venkatesan & Bashir 2011, Ayub & Bayley 2016, Howorka 2017, Shi et al. 2017, Long et al. 2018, Lee et al. 2018), however relatively muted has been an overview of the electronic signal-conditioning instrumentation to which such sensors are, or may be, interfaced. But such instrumentation is a critical component of the nanopore sensing system since the perceived limitations of such circuits (e.g. in throughput, sensitivity, speed, etc.) are often rightly cited as motivation for the introduction of new sensory innovations exploiting ever more exotic combinations of material arrangements and sensing modalities (e.g. tunnelling currents in place of ion blockades). Of particular interest is the state of microelectronic readout solutions implemented entirely in CMOS technology. CMOS-focused readout implementations are vital for they hold the means by which the most advanced nanopore-based sensing systems, in the form of biochips, may be realized. Conventional electrochemical sensors are designed to detect different events such as DNA hybridization, protein binding and other chemical / biological reactions. The aforementioned events are usually static, meaning that once the event takes place it will remain in the new state for a sufficiently long time. The required bandwidth for such configurations is usually low. Nanopore electrochemical cells on the other hand tend to have a more dynamic behavior. The events that are produced from an electrochemical cell nanopore tends to be very short in time mainly because these events are produced as a molecular structure or chain is being actually traversed through the pore as the measurements are commenced. Although there are techniques which reduce the molecular chain traverse speed however that would increase the chain sequencing time. Thus eventually with nanopore electrochemical cells the signal bandwidth is much higher than conventional electrochemical cells and the associated detection electronics will experience a higher in-band noise. As it will be shown in this review when converting the current signal of the nanopore electrochemical cells to a voltage equivalent, the voltage levels are relatively low due to the short available acquisition time. The acquired voltage in these devices can be comparable with random noise and crosstalk noise terms. Thus due to the aforementioned limitation conventional electrochemical cell readout electronics are not suitable for nanopore signal detection. The specifications of nanopore cells limit the performance of previously presented readout circuits used in conventional electrochemical cell sensors. Although both single channel and array based electrochemical sensors have been fully reviewed in previous studies, however as explained, nanopore electrochemical and solid state cells have specific characteristics such as low voltage/current swings, high bandwidth and high noise levels. With this in mind, CMOS readout systems and methods capable of addressing the challenges of state-of-the-art nanopore-based molecular sensors are reviewed. The focus will be on circuits suitable for nanopores employing ionic current sensing as this continues to be the most dominant and commercially relevant means of nanopore sensing. Ionic-current methods also have a particularly pressing need for effective readout systems given the aggressive channel speeds and measurement throughputs targeted by such methods.

To motivate this and to provide the reader some immediate context within the review, a discussion of the dominant nanopore sensing modalities is begun in Section 2 (which includes references to more in-depth coverage for extended exploration) and include a quick outline of the state of nanopore sensing structures (for ionic current measurement) in Section 3. An overview of continuous- time (Section 4.1) and discrete-time CMOS readout front-end components (Section 4.2) is then presented. This is followed by a discussion of the possible arrangement of these in high-throughput readout array systems (Section 5) that span nanopore as well as related electrochemical readout methods. A summary and closing discussion appear in Section 6.

## 2 Nanopore Sensing Methods

Nanopore-based particle sensing is currently being pursued along several main lines of study: resistive pulse shaping (RPS), transverse current detection, optics, and plasmonics. These avenues are briefly outlined below, emphasizing the advantages still available to the more mature RPS approach.

### 2.1 Resistive Pulse Shaping

The RPS sensing method is conceptually the simplest means of using nanopores to elucidate molecular structure. Perhaps for this reason it was the first nanopore sensing modality to achieve a relevant commercial impact (Clarke et al. 2009). Nanopore RPS works by immersing the nanopore sensor in an electrolyte solution across which a DC voltage $V_{dc}$ is applied. This potential prompts the solution's ions to flow through the sensor, as illustrated above in Figure 1, resulting in a small baseline DC current

$$I_{base} = G_p V_{dc} \qquad (1)$$

where $G_p = G_{dc}$ is the equivalent conductance of the pore under DC conditions. $I_{base}$ and $G_p$ depend on the nanopore's construction, $V_{dc}$, and ionic solution with typical molecular sensors (ranging across biological and solid-state nanopore constructions) operating between about 0.1 and 30 nA (Shekar et al. 2016) and $1/G_p = R_p$ between about 0.1 to 5 GΩ. Detailed models for $G_p$ (as well as the conductance blockade $\Delta G_t$ noted below) and the physics motivating them are discussed in Kowalczyk et al. (2011), Willmott & Smith (2012), Kowalczyk et al. (2012).

When a molecule translocates through the pore it impedes the baseline flow of ions through the sensor, reducing the conductance of the pore by $\Delta G_t$ in some relation to the molecule's fine structure (again, Figure 1 conveys this in an abstract manner), thus resulting in

$$G_p = G_{dc} - \Delta G_t. \qquad (2)$$

During these periods of occlusion, the current through the pore becomes

$$I_p = I_{base} - \Delta I_t = I_{base} - \Delta G_t V_{dc} \qquad (3)$$

where $\Delta I_t$ is the translocation blockade signal, the perturbation used to infer the blocking molecule's structure. Relative blockades, $\Delta I_t/I_{base}$, around 0.5 are typically observed in DNA experiments (Venta et al. 2013).

Ideally, the smaller the nanopore, the more accurately the RPS method can be used to resolve the structure of translocating molecules. Practical pores have facilitated this by achieving diameters as small as 2 nm. However, the reported thickness, $t$, of these openings has tended to be larger and the prospect of reducing it below the critical feature spacing in complex molecules such as DNA (e.g. ~0.34-nm) is daunting. Typical commercial-grade nanopore thickness may be on the order of 10 nm which means that in an RPS scheme as many as 30 bases may be simultaneously contributing to the blockade current. Having this many simultaneous contributors to the blockade signal is a clear challenge to signal fidelity and downstream signal processing.

A straightforward attempt to address this problem, that of increasing the signal amplitude (and thus the ability of sophisticated detectors to discern complex patterns) via a boost in $V_{dc}$ and hence $\Delta I_t$, is frustrated by the associated increase in DNA translocation rate, $f_t$, measured in base-pairs per second (bp/s). Although the signal's amplitude grows its broader spectral profile forces the readout circuitry to accumulate more noise as well.

Denoising methods and algorithms also play a critical role in the ultimately achieved SNR. Shekar et al. (2019) showed that wavelet denoising can be much superior in signal denoising of nanopore waveforms compared with the conventional Bessel approach.

Several different biological and solid state (including $MoS_2$ and SiNx) nanopores are studied regarding their RPS signal to noise performance by Fragasso et al. (2020). It is observed that SiNx solid state nanopores provide the highest signal to noise ratio under native translocation speeds. However, since biological nanopores can be equipped with the means to slow down the translocation, it is shown that ultimately biological nanopores which benefit from this advantage can achieve signal to noise ratios necessary for actual DNA sequencing. In the work, it is shown that the controlled slowdown of the translocation speed can be considerably effective in the obtained signal to noise ratio.

To this day the detection of ionic currents through biological nanopores based on the RPS approach still remains the dominant solution for sequencing applications, though research conducted to improve the performance still continues. Hartel et al. (2019, 2018) showed that the detection electronic plays an important role in increasing the temporal resolution. It is shown that microsecond temporal resolution is achievable using nanopores directly integrated on high-bandwidth, low-noise CMOS trans-impedance amplifiers.

## 2.2 Transverse Currents

Inspired by the aforementioned issues with nanopore RPS, researchers have proposed and pursued alternate nanopore-based excitation strategies, one being a lateral excitation apparatus. Specifically, by embedding electrodes at opposite sides of a nanopore opening, it was suggested that a finequantum tunnelling current could be made to flow across the pore (Zwolak & Di Ventra 2005), perpendicular to its opening. In theory, any ensuing disruption of this transverse current by moleculecomponents moving through the hole could be used to identify the translocating particle in greatdetail. For instance, electronic transport simulations assuming a 1.5-nm tunnelling gap betweengraphene electrodes biased at 1 V predicted mean current fluctuations on the order of $10^{-4}$, $10^{-5}$, $10^{-2}$, $10^{-6}$ pA for bases A, C, G, T, respectively (Prasongkit et al. 2011). An excellent recentreview of this approach is provided in Ventra & Taniguchi (2016). This approach also benefits bydecoupling the force that drives the molecule through the pore, $V_{dc}$ that is, from the measuredsignal. Thus may the translocation of the molecule to be slowed without reducing the size of thesignal indicative of the molecule's structure.

Experimental efforts in this direction have resulted in the ability to elicit signals on the order of 100 pA (albeit with noise fluctuations on the same order) from individual (dissolved in a fluid) DNA bases (Tsutsui et al. 2010, Huang et al. 2010) thus strongly demonstrating the capability of this technology for DNA sequencing. Efforts to measure contiguous polymer chains were also successful, although unable to discern the underlying structure of the molecule (Ivanov et al. 2011). Until now, this method has not met with the success achieved by RPS although this may also be a reflection of the great degree to which industry has attempted to solve the myriad of complications associated with practical RPS nanopore systems.

Along similar lines, researchers have been keen to exploit nanowire-like field-effect transistor (FET) measurement methods, these having previously been demonstrated for sensitive real-time chemical detection (Cui et al. 2001). Such technologies themselves have been inspired by the famous ion-sensitive FET (ISFET) approach (Bergveld 1970), and, besides improved signal quality, hold the potential for achieving extremely dense sensor arrays. In analogy to the tunnelling modality described above, FET-nanopore devices also rely on the modulation of a transverse current as an indicator of molecular make-up. Structurally, nanopore FETs consist of a semi-conductive material,a nano-channel, in close proximity to (or surrounding) a nanopore. A current is established throughthis conductor, but instead of molecularly modulating it via the re-arrangement of quantum states(as with conventional tunnelling methods), rather the expectation is that charge on the molecule is screened by charge in the nano-channel. As the molecule translocates the pore, in close proximityto the nano-channel, this screening should theoretically modulate the transverse current in waysunique to the passing molecule's make-up (Nelson et al. 2010). In this regard, the application of nanopores has not been limited to DNA sequencing and functionalized nanopores have been shown to be effective in high speed single molecule detection of proteins. FET equipped, aptamer functionalized nanopores have been used to detect human thrombin protein by Ren et al. (2020). However, ensuing experimental investigations implied a different physics at work in these devices (Xie et al. 2012, Traversi et al. 2013) than that outlined above. Likely due to the highly conductive ionic solutions, it was estimated that, in place of variations in the local density of states driving modulations in the transverse current, changes in the local potential of the fluid itself (caused by molecule translocation through the pore) were sensed by the nano-channel. Although this approach relies on a signal derived from the contribution of the molecule as a whole (or at least some subset of it), and thus undercuts somewhat an initial motivation for the use of transverse current modalities (i.e. finer resolution of molecular structure), it still holds excellent potential for improved signal and bandwidth performance over RPS methods (Parkin & Drndić 2018).

In this regard, the application of nanopores has not been limited to DNA sequencing and functionalized nanopores have been shown to be effective in high speed single molecule detection of proteins. FET equipped, aptamer functionalized nanopores have been used to detect human thrombin protein by Ren et al. (2020)

Traverse current measurement has also been made possible with nanoribbons. Since nanoribbons have single atom thickness, they have the potential to increase detection resolution by sensing one nucleotide at a time. Also it is shown that the achieved traverse current has much larger current changes compared with the ionic current approach (Heerema et al. 2018).

Using molecular dynamics simulations, combinational methods which process longitudinal and traverse currents have also been shown to be promising in enhancing the ability to distinguish between oligonucleotides during fast translocations (Farshad & Rasaiah 2020).

## 2.3 Optics and Plasmonics

As applied to DNA sequencing, optical methods based on polymerase-assisted addition of fluorescently labelled nucleotides are presently the dominant means of molecular measurement (Mardis 2017, Shendure et al. 2017). In the commercial market, coupled with the *sequencing-by-synthesis* (SBS) approach, these optical methods appreciably exceed the raw accuracy currently demonstrated by nanopore-based sequencers. This benefit stems from the superior signal-to-noise ratio(SNR) afforded by SBS at the cost of real-time functionality and direct molecular measurements (i.e. PCR-based molecular amplification is required excluding the ability to directly measure unaltered source molecules).

Workers have sought to port this sensing modality to nanopore platforms as well without compromising the single-molecule-detection and real-time processing advantages of deployed RPS nanopore solutions. At the moment the added benefits of this approach seem marginal; unlike purely current-based methods, proposed optical techniques require substantial tampering with the molecular analyte (e.g. expanding each nucleotide into multiple copies) to make room for fluorescent probes intended to work in a nanopore context (Soni & Meller 2007). However, such structural sequence dilations would effectively address the resolution challenges of RPS approaches as well and without the need to resort to bulky optical hardware (mainly for fluorescent probe excitation by laser). Nonetheless this optical method has been demonstrated (McNally et al. 2010) and may serve as a useful complementary means of measurement to help deal with the accuracy challenge currently faced by RPS.

The presumed ability of optical methods to simultaneously address multiple sensors has been noted as another of its benefits and has motivated a label-free optical sensing approach based on $Ca^{2+}$ induced photonic intensity (Huang et al. 2015). Although the opportunity to map a large array of nanopore-molecule reactions onto an imager is interesting, its advantage over pure electronic readout (as in RPS) is not obvious given the present and evolving technological landscape; ultimately all readout methods face the need to separately address individual or small groups of measurement sites (due to limits on the number of available pinouts from a chip). And although established SBS methods require optical pixel areas roughly 100× smaller than those needed for RPS systems, the signals they receive may originate from 1,000 copies of the same nucleotide, real-time nanopore-based sequencing methods, as single-molecule sensors, do not accommodate such redundancies.

Of qualitative similarity to the tunnelling current method are plasmonic nanoantennas which focus laser light into sub-diffraction limited "hotspots" that may be monitored by an imager. Locating such a hotspot across a nanopore creates the conditions under which a translocating molecule disrupts the highly focused light in ways indicative of its structure. Similar to the tunnelling current approach, the (label-free) plasmonic method has been demonstrated for macromolecule detection, i.e. the presence of a DNA macromolecule is registered, but not down to individual base level (Verschueren et al. 2018). As with the ideal transverse current approaches, this method decouples the forces driving molecules through the nanopore from the signals indicative of their structure. It also promises higher bandwidth operation, i.e. bandwidths around 5-kHz have been demonstrated (Shi et al. 2018), but its advantage over nanopore-FET methods which are more amenable to integration and possess estimated bandwidths around 100-MHz (Parkin & Drndić 2018) is not yet certain.

Optical methods based on spectroscopy and nanoslits have been also used for high spatial resolution sequencing of DNA chains that are threaded through nanopores (Chen et al. 2018).

## 3 Nanopore Sensor Structures

Since nanopore sensor development continues unabated, in this section performance-leading methods and emerging techniques in the area are described as pertains to ionic current measurement approaches.

### 3.1 Biological Nanopores

Biological nanopores (biopores) were the first nano-apertures to demonstrate the ability to detect molecular structure. Despite intensive efforts to develop solid-state embodiments, biopores have been the subject of the most concerted industrial attention and, perhaps consequently, have achieved the most compelling successes with regards to DNA sequencing (e.g. the assembly of a contiguous human genome (Jain et al. 2018)).

Among biopores, protein-based designs are the dominant motif. Using genetically engineered templates biotechnologists tune natural aperture-bearing protein structures into devices well suited for molecular sensing. Common alterations include changes to the amino acid content of a protein wild-type to affect dimensions, charge, hydrophobicity, obstructions, etc. Although not the focus of this review, it is worth noting that alternative biopore construction methods exist including those employing peptide assembly, DNA pores, and synthetic organic channels (Howorka 2017).

Besides now common biopore structures like alpha-hemolysin ($\alpha$HL) and *Mycobacterium smegmatis* porin A (MspA) (Song et al. 1996, Derrington et al. 2010), leading DNA strand-sequencing technologies have recently adopted variations on the *E. coli* amyloid secretion channel CsgG (Figure 2a) (Goyal et al. 2014). Noted advantages of this structure include a small minimum constriction diameter (0.9 nm adjustable to 1.5 nm) along with a wide channel opening (about 5 nm compared to 3 nm for $\alpha$HL) capable of better accommodating additional enzymes; a short distance between channel opening and constriction (4 nm compared to MspA's 9.8 nm); and relatively small pore thickness (10.5 nm compared to ClyA's 14.5 nm) (Howorka et al. 2017). Although longer range influences persist, the relatively small thickness of the CsgG's constriction (about 2 nm) ensures that only about 5 nucleic acid bases make the largest contribution to ionic current modulation. Under standard electrolyte conditions the pore exhibits a $R_p$ around 1 G$\Omega$.

Emerging developments in nanopore-based DNA strand-sequencers include the introduction of pores with multiple recognition sites (Stoddart et al. 2010), that is, sensors with more than one appreciable constriction within its lumen. Such sensors achieve a form of time-diversity and thus may be expected to exhibit higher SNR as has traditionally been exploited in communications systems. Accommodating structures for such applications include the protein excreting channels (secretins) common to Gram-negative bacteria such as GspD (a modified example of which is shown Figure 2b). In this case, the naturally occurring structure possesses a periplasmic (central) gate (on the *cis* side of an ostensible nanopore sensing apparatus) and an extracellular (cap) gate (on the *trans* side). Although these designs possess relatively large constrictions for efficient protein translocation, mutated structures may be realized for DNA strand sequencing applications (Jayasinghe, Wallace & Singh 2018). Along similar lines, composite nanopore sensors comprised of two back-to-back CsgG pores have also been discussed (Jayasinghe, Wallace, Singh, Hambley, Jordan & Remaut 2018).

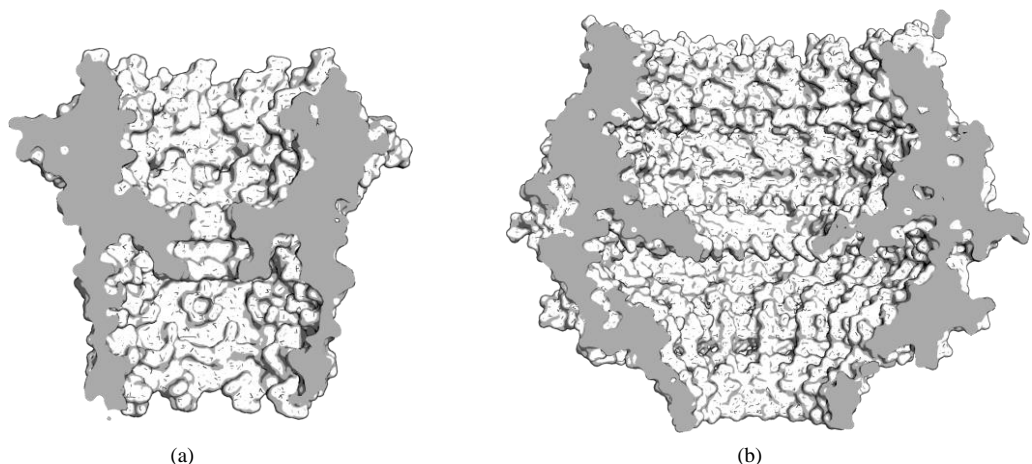

(a)           (b)

Figure 2: Examples of biological nanopore designs. (a) CsgG nanopore. (b) Altered GspD nanopore.

## 3.2 Solid-State Nanopores

Solid-state nanopores (SSNs) have the potential to demonstrate many advantages over biopore structures, these include not only the ability to operate across a broader range of environments (i.e. chemical and electrical), but also their promise to support a greater number of sensory modalities. Another related benefit often raised in SSNs' favour is the higher baseline current they can accommodate (e.g. $I_{base}$ exceeding 10 nA) compared to their biopore counterparts (which possess $I_{base}\sim$0.1 nA). This comes as a result of the lower thickness, $t$, and higher bias, $V_{dc}$, applicable to SSNs. However, unlike SSNs, biopores have been fit with extra molecular machinery (enzyme ratchets) to regulate their DNA translocation rates. As a result, biopores can offer output signal bandwidths (BWs), $f_t$, below 1 kHz whereas SSNs may easily exceed 1 MHz (Fologea et al. 2005). Thus, from a spectral density perspective, SSN and biopore signal integrity remain on a roughly equal footing. Readily accessible means of regulating SSN translocation rates, for example reducing salt concentrations, simultaneously compromise the available signal levels. In response, a number of authors have sought active means of molecule control through SSNs (Pud et al. 2016, Cadinu et al. 2018, Liu et al. 2018) with the ability of slowing translocation rates by two orders-of-magnitude without sacrificing SNR.

Judged by minimum achieved dimensions, among the most successful approaches to SSN formation have employed successive silicon nitride thinning steps, with the nitride being supported on a patterned $SiO_2$/Si substrate (Rodriguez-Manzo et al. 2015). Starting with a $SiN_x$ layer about 50-nm thick, the nitride is progressively thinned (the process takes approximately 30 minutes) across a small area (roughly 50×50 $nm^2$) to <2 nm diameter using repeated exposures to a scanning transmission electron microscope (STEM) probe. A final exposure in one spot achieves cylindrical apertures ~2 nm in diameter.

Although a highly controlled process, the STEM-mediated nanopore formation approach is challenging to scale to nanopore arrays (e.g. arrays of over 2,000 biopores within thumbnail footprints are now commercially available, achieving the same with one STEM probe would require over a month of fabrication time). Recent successes in replacing focused electron-beam-based fabrication with laser-based thinning are emerging as possibly more practical alternatives (Gilboa et al. 2018, Yamazaki et al. 2018). A potentially even more scalable approach to nanopore formation may be available via the dielectric breakdown approach (Kwok et al. 2014) by which randomly distributed nanopores as small as 2-nm have been formed in $SiN_x$ 10–30-nm thick via the application of a large electric field (<1 V/nm). Controlled means of implementing this technique using the tip of an atomic force microscope (AFM) have enabled the controlled formation of 300 pores in only 30 min (Zhang et al. 2019).

Besides achieving nanopore formation via the thinning of bulk substrates, there has been an energetic effort to incorporate 2D materials such as graphene and transition metal dichalcogenides (e.g. $WS_2$) in the nanopore construction process (Merchant et al. 2010, Schneider et al. 2010, Garaj et al. 2010, Heerema & Dekker 2016). Given thicknesses as small as 0.3 nm, such constructs are pursued for their potential to directly resolve individual bases (Feng et al. 2015).

As with the methods above, the current practice of using focused ion/electron beam techniques to perforate these materials presently limits their use in large sensor arrays.

Solid state nanopores provide the opportunity to reduce the pore thickness and thus decrease the number of nuceotides passing through the pore at any given time and increasing resolution.

Recently, monolayer materials such as $MoS_2$ have proven very useful for this purpose and differentiating single nucleotides. In this approach traverse currents are also used in the signal detectionand acquisition (Graf et al. 2018).

FET devices are also used to detect traverse electric fields and currents in nanopores and can be an effective solution in increasing the signal to noise ratio of the detected events. Wen & Zhang (2020) showed that since at high frequencies the external amplifier parasitic input capacitance is a dominant source of noise, a nanopore cell equipped with a local FET can outperform conventional ionic current measurement devices.

## 4 CMOS Readout Front-Ends

The apparatus by which nanopore-sensor electrochemical output signals are readied (i.e. amplified, filtered, digitized) for ensuing digital signal processing (DSP) is the *readout* block. Readout components have been a staple of biophysics and electrophysiology labs for decades where, for example, they support investigations into cellular ion-channel properties. Recent industrial developments have been successful in porting the extensive collections of off-the-shelf equipment traditionally used for readout into a microelectronic instantiation, a readout integrated circuit (ROIC). The creation of such technology has had a profound impact on the applicability of nanopores to molecular sequencing for sophisticated life-science practitioners, not to mention the extreme reduction in size afforded by an IC implementation. A key contribution of these ROIC developments involves their ability to fully leverage the power of semiconductor implementations by arraying the readout elements across a chip, thus realizing a high-throughput device. This multi-channel aspectis reviewed in Section 5. Presently however, developments associated with the key CMOS circuit building blocks needed within each channel will be focus on, paying particular attention to the analog front-end (AFE).

Recent advances in CMOS front-ends directly applied to, or directly relevant to, nanopore sensors have addressed current-sensing circuitry appropriate for RPS, the measurement modalityupon which this review is mainly focused. All of these systems are an outgrowth of the classic patch-clamp architecture, a circuit by which a stable potential is applied across an ion-channel-infused membrane. In simultaneity with this voltage control, the patch-clamp draws the channel'sion current into one of its biasing nodes and converts it to a voltage (*I-V*). This is sometimesalso classified as a single-electrode voltage-clamp. Ultimately, these designs are fundamentally centred on transimpedance amplifier (TIA) circuit blocks. Standard TIA arrangements to find usein nanopore systems are shown in Figure 3.

The resistive-feedback TIA of Figure 3a is the most straightforward approach, an input current $I_{in}$ is funnelled through the resistor $R_F$ and thus produces a proportional voltage, $V_{out}$ (the feedback capacitance, $C_F$ may be present as a combination of parasitic and intentionally-designed contributions), but this is impractical for modern nanopore applications. For example, with signal waveforms centred around roughly 100 pA a TIA-feedback resistor, $R_F$, on the order of 1 GΩ would be needed to provide an output voltage of adequate amplitude. With typical CMOS technologies possessing materials affording about 1 kΩ/$\mu$m$^2$, resistive components of this size would require untenable IC footprints. Also, with parasitic capacitance at roughly 20 fF/MΩ, the bandwidth of such systems would not even reach 100 Hz. Thus a tradeoff immediately presents itself: increase $R_F$ to establish a certain *I-V* conversion gain, but simultaneously sacrifice the bandwidth over which such gain may be realized.

The implementation of such frontends in the nanometer processes such as the low power amplifier given by Dawji et al. (2019) brings about the opportunity to integrate the analog sections of the circuit with local analog to digital converters and even compact digital processors.

Clearly, nanopore-based signal conditioning imposes a significant circuit design challenge right from the on-set. These have traditionally been handled from the continuous-time (CT) perspective, but as the search for CMOS-based solutions has intensified, a growing number of contributions have considered a discrete-time (DT) approach. Both are reviewed in the following.

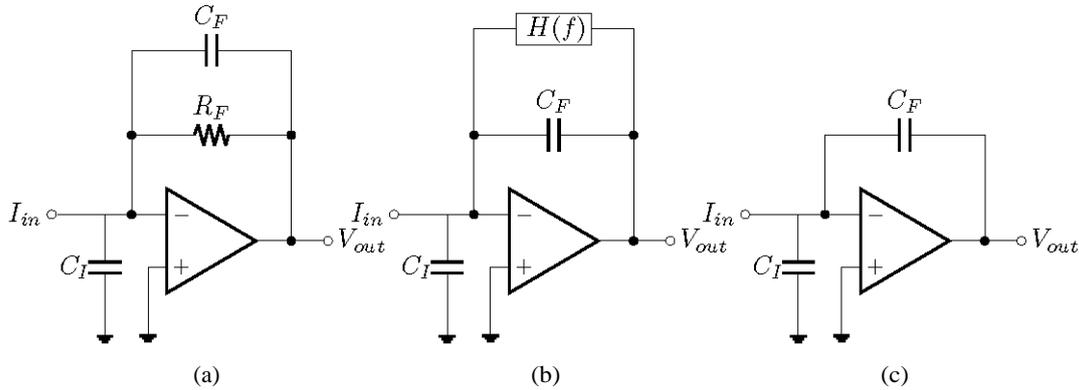

Figure 3: Transimpedance amplifier (TIA) main topology arrangements for current-based sensor readout. (a) resistive feedback TIA (b) CT capacitive feedback TIA (c) DT capacitive feedback TIA.

## 4.1 Continuous-Time CMOS Readout

A critical influence on the choices driving TIA design for nanopore applications is the input capacitance $C_I$ (shown in all parts of Figure 3). This component is due to not only the electronic elements of which the TIA is comprised but also depends on the sensory apparatus itself (e.g. the nanopore membrane/support structure, wiring connections, etc.). Even for state-of-the-art microfluidic nanopore-based sequencers $C_I$ can presently approach 5 pF.

An excellent feedforward TIA amplifier can marginalize this contribution, its high gain-bandwidth product (GBP) preventing any significant signal shunting through $C_I$ and thus resulting in a high bandwidth to seemingly affect high-speed nanopore-based DNA sequencing. Of course in this idealistic case, the bandwidth would be set by the $R_F C_F$ time constant, and thus emphasize for the designer the need for means of minimizing $C_F$. However a realistic TIA's performance is influenced by the mingling of $C_I$, $C_F$, and the amplifier's finite GBP; the introduction of a feedforward pole (at $R_F(C_I+C_F)$) and a feedback zero (at $R_F C_F$) combine to undermine stability at higher bandwidth unless impractically low values are used for $C_F$ (Huang et al. 2016).

As a result, the speed/stability clash encountered by nanopore-suitable (i.e. patch-clamp applications historically speaking) TIA implementations has traditionally been addressed by targeting a "well-defined frequency response" for the TIA (Sigworth 1995). This approach purposely augments $C_F$ to values that significantly separate the TIA's two main closed-loop poles despite a potentially high $C_I$ or less aggressive GBP. The most aggressive version of this method replaces the $R_F$ entirely with $C_F$ resulting in a pure integrator whose attenuation at increasing frequencies is corrected by an ensuing differentiator block. Among the advantages of this approach is that it makes the pole/zero compensation of the integrator/differentiator chain easier. However, this method clearly presents a problem for CT approaches vis-a-vis the need to periodically discharge the $C_F$. It is, however, a natural avenue to follow in the DT case as discussed in Section 4.2.

A reset free, high-bandwidth (4-MHz), capacitive integrator-differentiator scheme is achieved in 0.35 $\mu$m CMOS by implementing a special DC feedback loop (Ferrari et al. 2007, Ferrari, Gozzini, Molari & Sampietro 2009). Using a frequency dependent active feedback ($H(s)$ in Figure 3b) the DC current is removed from the input to avoid saturating $C_F$. This active feedback provides low attenuation at low frequencies (below 100 Hz) and high attenuation at high frequencies (so it does not compromise the integrator output). A high resistance is required to implement a low-noise DC feedback. A current-reduction circuit working in tandem with the frequency-dependent feedback essentially scales up a 300 k$\Omega$ on-chip reference resistance by six orders of magnitude.

An alternative continuous-time ultra-low-noise pre-amplifier uses a matched double-MOS scheme to multiply the input current (Ferrari, Farina, Guagliardo, Carminati & Sampietro 2009). Specifically, in place of $R_F$ the TIA uses a NMOS/PMOS pair (a *double*); at its output, the $N \times$ scaled circuit replica (with matching bias) of the double is present resulting in a translinear current multiplication by a factor of $N$ (ratios of 99 and 10 are tested). This amplifier can reach a noise floor of 500 aA/$\sqrt{Hz}$ under low input current. But as the input current increases the noise increases linearly due to the shot noise of subthreshold feedback transistors. Moreover, this circuit does not implement any flicker noise capture techniques. As a result, one needs to account for the flicker noise from the nanopore itself.

Another integrator-differentiator design built specifically to accommodate the high-frequency requirements of SSNs are discussed by Rosenstein et al. (2012, 2013). The integrator follows the general approach described by Ferrari, Gozzini, Molari & Sampietro (2009), but employs a much lower feedback resistance, $R_F = 100$ MΩ, and thus achieves its pole/zero cancellation at a much higher frequency (about 25 kHz vs. 100 Hz as noted above). From a signal-amplitude perspective the drop in $R_F$ is acceptable since SSNs accommodate larger currents. Similarly, the possible distortion invoked by the hike in the circuit's lower corner frequency due to the drop in $R_F$ (i.e. below $R_F C_F^{-1}$ the differentiator effectively distorts the signal) is manageable for the higher signal speeds inherent to SSNs (relative to ratcheted biopores). This design also uses a linear current source as an active feedback resistor for low noise. A significant improvement in area was also achieved relative to Ferrari, Gozzini, Molari & Sampietro (2009) as a complex DC feedback loop is avoided.

## 4.2  Discrete-Time CMOS Readout

As intimated above, DT ROICs may also follow the integrator-differentiator scheme albeit by embracing a purely capacitive feedback in the integrator stage along the lines shown in Figure 3c. An early CMOS example of such an approach for nanopore applications is discussed by Goldstein et al. (2012). This work presents a DT low-noise amplifier consisting of an integrator, a post amplifier and a low-pass filter. Correlated double sampling (CDS) is performed off-chip. Besides attenuating the offset and flicker noise, the CDS effectively realizes the differentiation function as well, but doubles the thermal noise. To address this latter point, Goldstein et al. (2012) relies on a low thermal noise amplifier in the integrator.

Contemporaneously, Lu & Holleman (2012) described an ultra-low-current measurement technique employing *capacitive matching*. Typically, an AFE for nanopore-based measurement consists of a low noise amplifier, a filter and an analog-to-digital converter (ADC). These blocks introduce signal distortions due to issues such as charge injection in the amplifier and analog imperfections in the ADC. To overcome these problems Lu & Holleman (2012) employs a nested auto-zero scheme and a novel CDS. A threshold comparator, with two threshold levels, is used to measure the slope of the integrator output. After the auto-zero phase, the threshold of the comparator is set to the first threshold level. Once the output the comparator goes high, its threshold is switched to the second threshold level. The time-difference between the two high outputs of the comparator, which is counted using a counter, infers the slope of the integrator. This mechanism acts like a conventional CDS reducing offset and flicker noise. Thus, the input current is measured as a function of pulse width. This eliminates the need for an ADC and marginalizes non-linearities caused by analog mismatch.

A mostly continuous-time circuit that includes re-setting out of concerns that the input amplifier may saturate due to accumulated charge on the input capacitance itself is presented by Kim, Maitra, Pedrotti & Dunbar (2013). A typical instrumentation approach is followed by using a resistive feedback TIA. This causes higher noise and occupies more area.

In the DT amplifiers, the transimpedance gain ($R_{EQ}$) of the integrator is propositional to the integration period ($T_{INT}$), $R_{EQ} = T_{INT}/C_F$. All conventional DT circuits have an equal reset rate ($f_r$) and sampling rate ($f_s$). To achieve higher bandwidth in conventional DT amplifiers, the integration time needs to be reduced. This in turn reduces the gain of the system and increases the input-referred noise, a major constraint of conventional DT amplifiers. Crescentini et al. (2017) propose a system where the integrator is re-set at a rate that differs from the ADCs sampling frequency. By sampling at the higher rate, a signal with higher bandwidth can be captured while the reset rate is kept constant. Thus, the gain and the input-referred noise remain the same. The authors proposed a track-and-hold (T&H) after the integrator-differentiator to avoid sampling the corrupted signal during the reset phase. Alternatively, this can also be done digitally. A low pass filter is also employed to filter out high frequency spurs caused by the T&H.

A completely novel approach using an hourglass ADC, which has similarities to a sigma-delta converter, has been presented by Hsu & Hall (2018). An hourglass ADC employs a capacitive TIA in feedback. To avoid saturation, a switch flips the polarity of the input signal within a user defined time window. This folds the output of the integrator, converting it from a ramp into a triangular wave and prevents it from saturating. This also converts the output of the integrator from a voltage to a frequency which provides high dynamic range to the hourglass ADC. To improve linearity, the input is also connected to a steering DAC which measures the coarse input current and reduces the input to the hourglass ADC.

Table 1: Nanopore-Relevant DT and CT AFE Performance Comparison

| Ref. | Style | Input Range [nA] | Noise Floor [fA/√Hz] | @kHz | BW [kHz] | Integ. Noise [pArms] | Inp. Cap. [pF] | Input Res. [GΩ] | Power [mW] |
|---|---|---|---|---|---|---|---|---|---|
| Ferrari et al. (2007) | CT | 25 | 4 | @100 | 0.1–1000 | – | 6 | 0 | – |
| Rosenstein et al. (2012, 2013) | CT | 10 | 10 | @10 | 0.1–100 | 3.2 | 6 | 0 | 5 |
|  |  |  |  |  | 0.1–1000 | 24 |  |  |  |
| Ferrari, Farina et al. (2009) | CT | 40 | 0.5 | @10 | 0–100 | 1 | 5 | 0 | 5 |
| Hsu et al. (2015) | CT | 10 | 8.5 | @10 | 100–10 | 3.4 | 10 | 1 | 65 |
|  |  |  |  |  | 0.1–100 | 13.5 |  |  |  |
| Taherzadeh-Sani et al. (2017) | CT | 10 | 11 | @10 | 0.1–10 | 0.4 | 6 | 10 | 30 |
|  |  |  |  |  | 0.1–100 | 3.4 |  |  |  |
|  |  |  |  |  | 0.1–1000 | 52 |  |  |  |
| Dai et al. (2016) | CT | 11 600 | 11.6 | @1 | 0–10 | 1.25 | 0 | 0 | 1.5 |
| Dawji et al. (2019) | CT | NA | 8.75 | @10 | 0–10 | 1.6 | 5 | 0 | 0.6 |
| Goldstein et al. (2012) | DT | 0.2 | 5 | @1 | 0–10 | 2.44 | 47 | 10 | 1.5 |
| Lu & Holleman (2012) | DT | 0.2 | 0.235 | @1 | – | – | 0 | – | 0.15 |
| Kim, Maitra, Pedrotti & Dunbar (2013) | DT | 0.05 | – |  | 0-10 | 8.3 | 50 | 1 | 3.8 |
| Crescentini et al. (2017) | DT | 0.2 | 4 |  | 0-10 | – | 3 | 1 | – |
| Hsu & Hall (2018) | DT | 10 000 | 30 | @1 | 0–1 | – | 0 | 0 | 0.29 |
| 200B | Cap. | – | 0.7 |  | 0–100 | – | 0 | 0 | – |
| 200B | Res. | – | 6 |  | 0–100 | – | 0 | 0 | – |

A summary performance comparison of a number of CT and DT designs discussed above is presented in Table 1. Owing to the myriad design approaches and even measurement (e.g. the input source capacitance/resistance and the DC level of the input signal) specific conclusions on performance need to be carefully tempered. Generally speaking, DT amplifiers have lower noise while CT amplifiers can amplify signals to a much higher bandwidth and have a higher dynamic range. Also included in Table 1 is a comparison of the CMOS proposals to common benchtop instrumentation systems (Axopatch 200B). Clearly, from a noise and bandwidth perspective the CMOS approaches offer highly competitive performance levels (not to mention their intrinsic benefits in cost and size).

## 5 CMOS Readout Arrays

Although the design and realization of high-performance readout components for biosensors remains a critical objective, their arrangement into high-throughput parallel-channel arrays is arguably even more important in today's rapidly evolving bio-technological landscape. The role of ISFET-based DNA sensing paints a particularly vivid example of this. The introduction of these devices in the early '70s as a transistor-based means of chemical analysis (Bergveld 1970) effectively launched research into bio-sensing interfaces to active substrates. The realization of these sensors within a commercial CMOS process (Bausells et al. 1999) 25 years later prompted, after another five years, realizations of the first four-element ISFET CMOS sensor arrays (Milgrew et al. 2003, 2004). Within four years of these demonstrations, by 2007, others reported 2D arrays consisting of 260k ISFET pixels (Rothberg, Hinz, Johnson & Bustillo 2011), by 2011 fully operational DNA sequencers based on 1.5M-pixel chips were announced (Rothberg et al. 2011); a per-pixel area of roughly 20 $\mu m^2$ in a 350-nm CMOS technology.

Judged by element count alone, the scale of integration achieved by ISFET-based CMOS biochips remains unparalleled in the field. This is, at least in part, realized by the extreme degree to which the complexity of ISFET sensing pixels may be reduced, primarily at the cost of dynamic range, to accommodate certain means of sequencing, i.e. broadly speaking, the SBS method (McCombie et al. 2018, Merriman & Rothberg 2012). For DNA sequencing, each ISFET pixel may be reduced to two or three transistors producing raw signals with magnitudes and bandwidths on the order of 10 mV and 0.1 Hz with a final SNR exceeding 20 dB at the detector (Rothberg et al. 2011). Being largely (although not exclusively) intended as single-molecule detectors, nanopore-based readout methods lack the advantages available to SBS systems which benefit from stronger signals due to redundant molecular samples and finely controlled chemical reactions (allowing for lower bandwidth operation).

As a result of their signal characteristics, nanopore-based measurement systems require more intricate readout components (as discussed in Section 4 above) and hence need more sophisticated front-ends within each pixel as elaborated in Section 5.2. Detailed reports on means of effectively arraying such pixels for high performance nanopore-based molecular measurement, specifically DNA sequencing, are practically non-existent in the academic literature; more common, in this sub-specialty, are descriptions of individual component design (again, as discussed in Section 4 above) and means of addressing the performance stress imposed upon them by challenging nanopore signal characteristics.

From an arrayed CMOS biochip perspective, much more active have been efforts amenable to nucleic acid testing (NAT), the detection of DNA oligos containing pre-determined sequence patterns. These technologies, an outgrowth of DNA microarrays, have prompted the design of a rich assortment of arrayed CMOS biosensing chips. The design motivations, challenges faced, and successes achieved by such realizations inform nanopore-based CMOS sequencing as well. Outside nucleic acid measurement, but along cognate system-design threads, are a number of other electrochemical techniques which have demonstrated insightful CMOS arrays. Thus, in the following not only emerging work in nanopore sequencing arrays are reviewed, but the contributions of their NAT-amenable and relevant electrochemical sensing counterparts are covered as well.

## 5.1 Hybridizing Reaction Electrochemical Sensor Arrays

Sensing via hybridization (bonding) is now an extremely common, hypothesis-based, means of screening for DNA patterns. In short, the sensing is based on observing chemical reaction products from interactions between matching DNA molecules; pre-made and physically localized 'probe' DNA and freely moving 'target' DNA whose identity is sought. If a reaction in a particular location is observed, the sequence of a target DNA molecule is immediately revealed. This approach, in the form of DNA microarrays, began to leverage semiconductor manufacturing methods (albeit on glass, not on semiconductor substrates) and optical detection in the early '90s (Fodor et al. 1993). Presently, such passive microarray technologies offer on the order of 500k sensor sites over dimensions of 1 $cm^2$.

Some important challenges, which are still being addressed by researchers, include means of increasing the array resolutions and cell sensitivity. With an improved cell sensitivity, hybridization detection can be performed at lower concentrations and with higher resolution the sequencing throughput can be increased. To increase sensitivity, other detection methods besides optical ones have been investigated; among these, electrochemical cell-based techniques have arguably had the biggest contribution (Jafari et al. 2014). These are based mainly on charge detection, current detection, or cell impedance measurements and the circuits used to process these modalities come closest to those used in nanopore AFEs.

As higher array resolutions are pursued, processing the large amount of data produced inside the chip becomes ever more challenging for its circuitry to deal with. Specifically, in recent detection methods that perform nucleotide chemical cycles on chip (Fuller et al. 2016), and thus present a new signal with each cycle, the real-time nature of the measurement scheme and high-speed detection of the events during the processing cycles becomes more important.

A big challenge in this regard is to forward an approach that can get the data out of the large cell array. The sooner the data can be extracted out of the array, the sooner the next oligonucleotide detection step can be commenced. A small time step for each nucleotide detection will naturally result in faster sequencing. Thus, depending on the application of the detection array, whether it is based on an optical approach or one that employs electrochemical cell electrical sensing, and depending on the type of event that happens on the hybridization surface, different solutions for processing and extracting the data have been proposed.

In a hypothesis-driven approach where a single event is to be detected (i.e. whether a particular DNA oligonucleotide pattern is present or not) without the need for further chemical cycles, a single ADC and/or AFE per entire array is sufficient to extract the produced results. For higher extraction speeds as encountered in DNA sequencing and in signal extraction under multiple chemical cycles, other front-end configurations such as column based and pixel based circuitry are more common (Rothberg et al. 2011). For short-sweep time-cyclic voltammetry electrochemical tests, the signal produced inside the array is instantaneous and fast detection of the event is necessary. In this case the use of column parallel or pixel based AFEs is also a more effective choice.

Most of the reported CMOS hybridization arrays that are capable of sensing chemical events have a conducting electrode or a photodiode as the sensing site. Other circuits which can be integrated inside each pixel include amplifiers, current mirrors, integrators, mixers, and ADC circuits. Integrators are commonly used to receive the electrode/photodiode current and convert it to a voltage signal. This procedure both converts the current to a voltage signal and increases the SNR due to the averaging effect of the integrator. In-pixel ADCs are used to convert the detected analog electrical signal to a digital representation so that it can be acquired with high access speeds in large arrays. Some work regarding this categorization is reviewed in the following.

The addition of scalable and parallel channels to the electrochemical sensor has always been a demanding need for bioelectrochemical systems. With the higher number of channels, the effective experiment throughput can be increased. A time multiplexed approach has been presented using 128 individual Linear Quadratic Integral (LQI) controllers by Molderez et al. (2021). Althoughthe method reduced implementation cost however since the given approach uses time multiplexingthe obtained bandwidth is not high enough for nanopore DNA sequencing.

Integrated CMOS electronics promises high density arrays of readout channel circuitry for electrochemical cell experiments. While high density arrays can be effective for imaging of pH activity and other slow response and low bandwidth events such as dissolved oxygen (Tedjo & Chen 2019), however large channel DNA sequencing using these arrays are relatively challenging due the high bandwidth and low noise requirements of the corresponding nanopore electrochemical cells.

Although high density arrays with high temporal resolution have been previously presented by Tedjo et al. (2018), however such arrays cannot perform simultaneous recordings of all channels at maximum bandwidth and the array has to be configured to distribute the available bandwidth among the available electrodes of the sensor.

### 5.1.1 Column-based and array-based front-ends

Among the first descriptions of a fully electronic hybridization-based CMOS DNA sensing array is offered by Thewes et al. (2002). Similar to established methods, the sensing procedure consists of labeled DNA *targets* which are capable of hybridizing with matching DNA *probe* molecules. This arrangement, traditionally reliant on optical sensing, is altered by Thewes et al. (2002) to enable purely electrical monitoring. After the target is introduced a chemical substrate is added and subsequently chemically altered upon reaction with target's label (an enzyme). This alteration causes negatively charged molecular mediators to be released. These charges are sensed as current contributors to a multi-electrode electrolytic cell undergoing redox recycling.

Using a 0.5-$\mu$m-technology, square pixels approximately 200 $\mu$m on a side are combined to achieve a 16×8 array within an area of about 3×1.5 mm$^2$ (a total chip area of 3.5×2.5 mm$^2$). Each pixel is relatively simple, containing two amplifiers to control the electrochemical cell potentials
and successive current mirrors operating in the subthreshold region to amplify and direct the electrochemical cell current. Via corresponding row and column select lines, all measured currents are multiplexed out to a single ADC which services all 128 channels. Since the current is amplified by the current mirror stages, the switching activity of the row and column buses has less effecton the sensor performance, however, with a single ADC, the scan rate of all channels is relativelylimited, with about 10 samples gathered over seconds or minutes.

Column ADCs are also used in the 200×120-pixel ion-sensing electrochemical cells presented by Jiang et al. (2018). The high cell density of this design has been achieved by using a very simple AFE architecture. The AFE integrator consists of an ISFET device and a capacitor as shown in Figure 4a. The ISFET acts as an amplifier, producing an output current in proportionto the electrochemical cell proton ions. The current then enters an in-pixel capacitor where it is integrated and converted to a voltage reading.

Kruppa et al. (2010) presents a 24×16 pixel DNA hybridization detection sensor. The attachment of a DNA specimen to the probe is detected using an electrochemical current measurement
at each of a number of specific bias potentials over a reference voltage sweep (a method called *cyclic voltammetry*). In-pixel, discrete current integrators, are used for this purpose to detect the instantaneous current produced during the sweep. In this work, a differential hybridization technique is used in which two probe sites exist for each cell. Upon hybridization one site receives the segment while the other site releases a pre-hybridized segment thus producing a differential signal; consequently, for the AFE, a differential integrator has been used. A single 12-bit SAR ADC is used for the entire 384-element array. The integration time for each cell is 6 ms. The circuit can measure currents as low as 100 pA.

The sensor presented by Rothe et al. (2011) consists of 24×24 electrochemical cell current-detection electrodes. Each column is connected to one individual sigma-delta potentiostat and
one ADC. Each column potentiostat and ADC process the currents produced by the electrodesof an entire column. The electrodes of each row are individually selected at each time frame and multiplexed to the corresponding potentiostat and ADC.

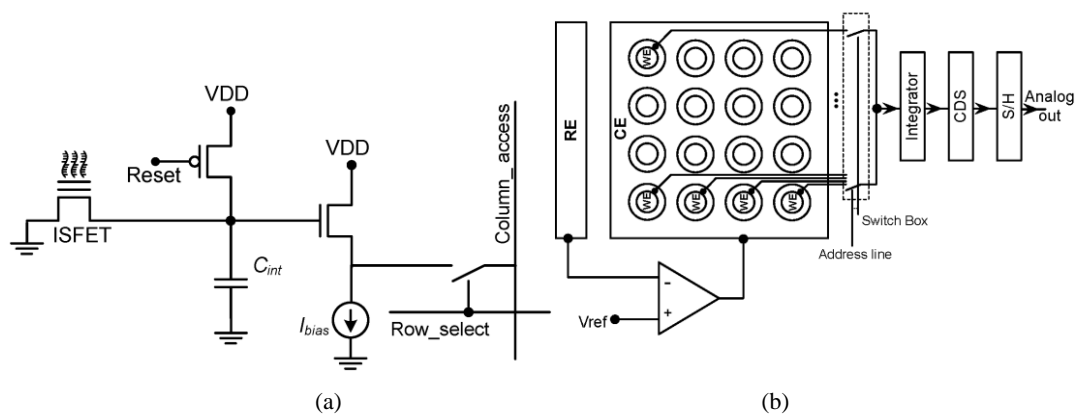

(a)          (b)

Figure 4: Examples of column-based and array-based front-ends. (a) An in-pixel integrator using ISFET devices as amplifiers. Column ADCs are used in this sensor to convert each cell's output voltage to digital (Jiang et al. 2018). (b) A block diagram of a passive electrochemical sensor array where each cell consists of only the sensing electrode (Yang et al. 2009).

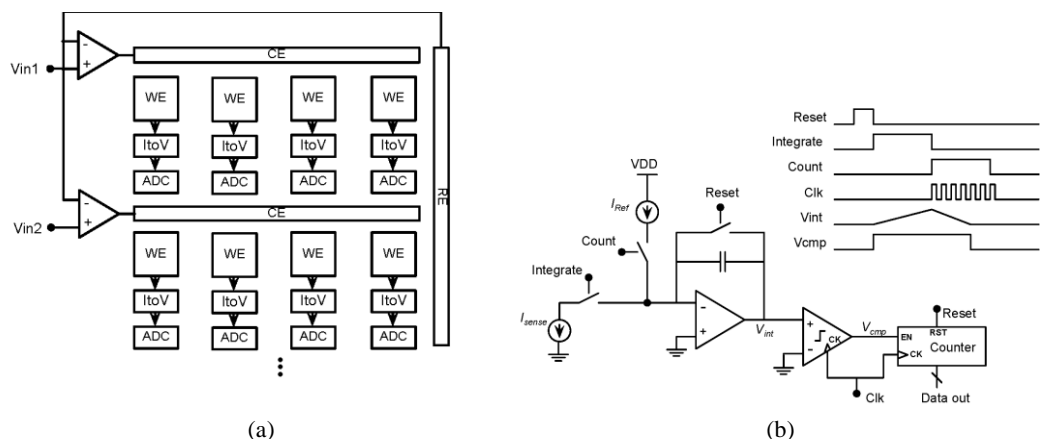

(a)          (b)

Figure 5: An example of an in-pixel-ADC array design (Levine et al. 2008). (a) Arrangement of sensing electrodes, AFEs and in-pixel ADCs in an electrochemical cell signal detection array. (b) Basic circuit schematic of the in-pixel dual-slope ADC and its corresponding waveforms.

Using redox reactors which have a slower response to voltage cycles it is possible to further reduce the voltage cycle sweep time and hence the sample rate. With a slower sample rate, a single ADC per array is more appropriate for the entire array. The 4×4 electrode array presented by Yang et al. (2009) has only one potentiostat and one ADC for the entire array and thus the sample rate of the entire array is relatively long compared to the previously proposed column based or pixel based processing sensors. The array configuration is shown in Figure 4b.

### 5.1.2 In-pixel ADCs

In-pixel ADCs are included in the electrochemical sensor presented by Schienle et al. (2004). Again, a 16×8 array was realized in 0.5-$\mu$m CMOS in a 6.4×4.5 mm$^2$ chip with square pixels roughly 300 $\mu$m on a side. In each pixel, a current conveyor amplifies the current and directs it to a 140-fF storage capacitor forming an integrator. Using a self-resetting method, the storage capacitor is reset whenever the voltage reaches a specific limit. Using an in-pixel counter the number of reset cycles per unit time, ranging from 7 Hz to 700 kHz for currents between 1 pA and 100-nA serves as a digital representation of the input current. As a result, to discern 1-pA fluctuations this system can process signals up to bandwidths on the order of 1 Hz.

The work by Levine et al. (2008) also presents a DNA hybridization detection sensor using in-pixel ADCs as shown in Figure 5a. The design is carried out using a 0.25-$\mu$m CMOS technology and achieves a 4×4 cell array in a 5×3-mm$^2$ chip and pixel dimensions of about 550 $\mu$m ×770 $\mu$m. This implementation detects the changes in the redox reactor concentration when the probe sites are empty or hybridized. Electrochemical cell current measurements are also used for this purpose with each pixel containing an ADC.

In-pixel dual-slope ADCs are used in this work as shown in Figure 5b. When the integrator's output signal reaches a specific threshold, the integrator capacitance is discharged with a known reference current, $I_{Ref}$. The amount of input current, $I_{sense}$, is then measured by evaluating the ratio of the integrator discharge time to the charge time. Using in-pixel timers (counters) a digital representation can be obtained inside each pixel. The conversion time of one sample is about 0.35 ms in this sensor.

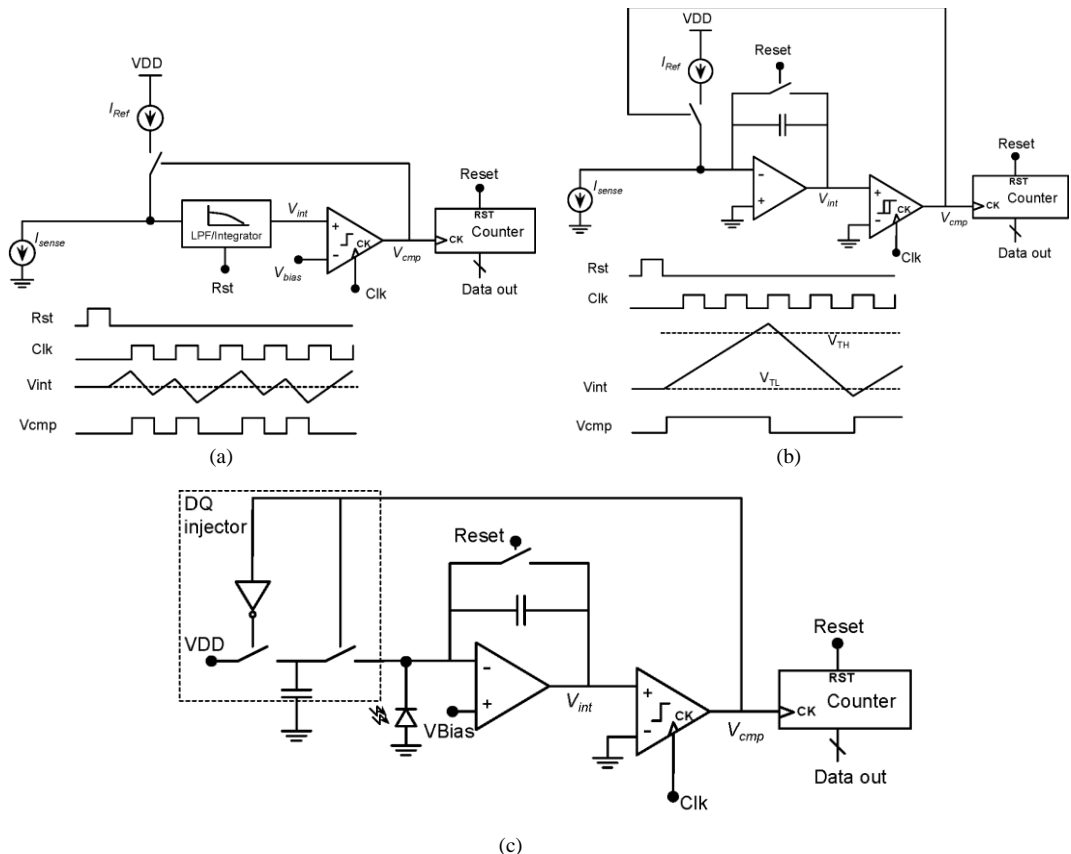

Figure 6: Examples of array designs employing in-pixel sigma-delta converters. (a) A simplified sigma-delta converter used as an in- pixel ADC. The control waveforms show at which sampling cycles a delta current is subtracted from the integrator (Sutula et al. 2014). (b) A modified sigma-delta converter with a hysteresis comparison band to reduce switching noise (Gore et al. 2006). (c) A sigma-delta converter used to convert photo diode current using charge delta packets. (Hassibi et al. 2018).

### 5.1.3 In-pixel sigma-delta converters

The sigma-delta approach is relatively effective as an in-pixel ADC due to its compact size. Compared to the in-pixel dual slope ADC, the sigma-delta ADC has better noise shaping characteristics due its inherently higher signal sampling rate. For the same reason, the sigma-delta ADC can better remove low frequency drifts and noise components such as flicker noise. A drawback of the sigma-delta converter is its higher switching noise which tends to perturb weak input signals with low voltage fluctuations. Equivalently, one could say the drawback of the sigma-delta converter is its high conversion time if the sampling frequency is reduced to alleviate the problem of switching noise.

Sutula et al. (2014), use a sigma-delta integrator as shown in Figure 6a for the in-pixel AFE. With the sigma-delta integrator, when the output reaches a particular reference value at the comparator, a certain amount is subtracted from the integrator. In the sigma-delta feedback loop, apiecewise waveform is generated at the comparator output (as also shown in Figure 6a). In this case, by counting the number of comparator pulses per unit time, a binary coded quantity can be obtained.

Gore et al. (2006) present a semi-asynchronous sigma-delta approach for a 42-channel potentiostat to reduce the switching activity of the converter at low input currents and thus decreaseswitching noise. As shown in Figure 6b, instead of using a single threshold comparator to decide when to inject the delta current packet, a two level hysteresis comparator is incorporated.

In-pixel sigma-delta converters have also been used in optical DNA hybridization detection. Hassibi et al. (2018) use a large array of pixels to detect DNA hybridization, however in this case colour labels are used in the hybridization detection process and hence photo detectors are realized inside each pixel. Similar to many electrochemical sensors with in-pixel ADCs, this detector also benefits from compact sigma-delta converters to obtain a digitized in-pixel result. However, as shown in Figure 6c instead of injecting delta current packets at the integrator input, constant charge packets are directed to the integrator using a switch-capacitor configuration. The proposed sensor also includes heaters to facilitate in-pixel PCR and amplification of the DNA fragments and specimens.

### 5.2 Resistive Pulse Shaping Sensor Arrays

Owing to its monomeric sensing and dynamic nature, the signal entering the nanopore front-end is relatively weak and sensitive to noise. Usually in all implemented front-ends a conventional integrator without sigma-delta switching activity is used. Since no noise shaping sigma-delta technique is used, usually a LPF is also required to filter out noise. Similar to the DNA hybridization sensors, nanopore cells are also used in array structures, however, due to the aforementioned challenges, currently the reported architectures have more complex in-pixel circuits compared to what is common in DNA hybridization detection array sensors.

The integration of the actual nanopore cell with the CMOS electronics helps eliminate external interferences and hence reduce noise. With lower noise levels higher translocation recordings are possible. A 10MHz bandwidth DNA translocation event recorder is presented which integrates glass chip nanopores with custom-designed CMOS amplifiers by Chien et al. (2019).

Current trends in electrochemical cell readout chips are to increase performance factors such as number of readout channels per chip, channel bandwidth and current precisions down to the pA range. An electrochemical readout chip is presented with 1.1pA accuracy and 100kHz bandwidth per channel by Li et al. (2019). The designed chip supports up to 64 readout channels.

The 5×5 pixel nanopore sensor presented by Anderson et al. (2008) has the integrator, gain amplifier stage, amplifier parameter adjustment logic and output amplifier all within a pixel. While each electrode occupies an area of $100\times100$ $\mu m^2$, each pixel with all the processing elements requires a die area of about $400\times400$ $\mu m^2$. The sensor does not include any filters or ADCs. The downstream processing functions, such as filtering and analog-to-digital conversion, are performed externally off-chip. After the data has been acquired and converted to digital at an oversampled rate the data is filtered in the digital domain.

Fish & White (2018) present a multichannel readout circuitry for nanopore arrays as shown in Figure 7. Multiple nanopores are connected to each channel and can be selected for signal processing. Each channel contains an integrator as the main amplifier and CDS circuits. The CDS functionality is achieved by sample-and-hold circuits which can sample the integrator output at the beginning and end of each integration period. In the work, it is described that, although active ADCs induce significant noise on the integrators, since their digital activity is AC in nature, when the ADC is halted the integrator output will return to its conventional reading. Thus, to eliminate digital noise issues a timing mechanism is proposed so that the ADC operation is only commenced after a sample is acquired by the sample and hold. In the time windows where the ADC is operational, the integrator output is isolated from the sample taken and thus the noise induced on the integrator input does not affect the ADC reading.

Conventional nanopore frontends need to have low noise and high bandwidth operational characteristics. Usually these specifications are obtained at the price of high power dissipation. In turn a large array of power hungry frontends can lead to thermal complexities with the DNA readout chip. Since as explained earlier, for low noise configurations, nanopores should be integrated with the actual CMOS readout circuitry the elevated heat can even compromise the actual DNA transfer through the pore. Thus suitable methods to reduce the power usage of the frontend channels can be relatively effective in the implementation of large nanopore cell and readout arrays. Such low power frontends can be found in multifunctional automated nanopore sequencing CMOS integrated circuit presented by Dong et al. (2020).

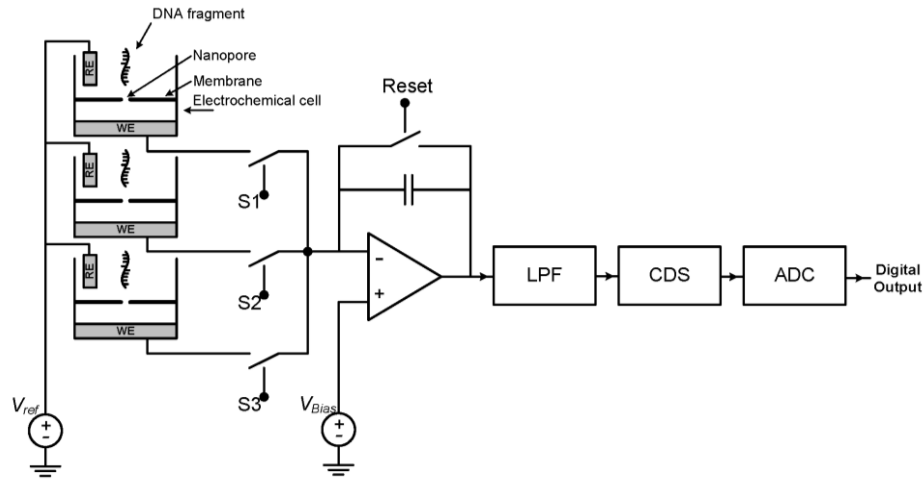

Figure 7: A conventional signal processing chain for nanopore current reception and conversion (Fish & White 2018).

As evident from available work on nanopore sensors, implementation of large arrays with compact in-pixel ADCs (of sigma-delta or dual-slope types for example) have been lacking in the literature. This is mainly due to the digital noise level and long conversion time of these types of in-pixel circuits. Another reason for a lack of such arrays is that nanopore front-ends are usually hardware intensive. For example, due to the integrator's own noise, a filter is usually required in the chain. While there have been different designs descriptions in the form of patent claims which suggest solutions that meets both the area limitations and nanopore interface requirements, there have been no measurement results confirming the effectiveness of such proposed approaches and practical solutions effectively remain as trade secrets.

One such design can be found by Hovis et al. (2016) where a simple and compact in-pixel structure is presented. As can be seen Figure 8a, with the proposed open-loop circuit, the electrochemicalcell is pre-charged to a certain potential and allowed to discharge. The voltage decay of the cell is then monitored by an amplifier. The signal is subsequently converted to a digital representationusing an ADC. By processing the decay time of the cell the DNA structure is analyzed.

Also Tian (2015) present a nanopore array with column based ADCs. Each column ADC is designated to service multiple columns through multiplexing. As shown in Figure 8b each pixel of the sensor is equipped with compact current conveyor integrators. Although the pixel size is relatively small in this claim, higher noise is expected in this method compared to the integration based methods as due to the thermal noise of the current mirror itself (Kim, Goldstein, Tang, Sigworth & Culurciello 2013).

In another description (Chen 2013) a nanopore sensor array with in-pixel event generation ADCs is presented. The pixels shown in Figure 9 use an integrator to accumulate nanopore current. Whenthe integrator output voltage reaches a specific level, an event is generated and the nanopore is reset.Owing to the random nature of the events generated throughout the array, the digital activity onthe data lines can induce noise on the integrator. To solve this problem, appropriate metal shieldingtechniques are proposed in the work. The lines that generate the events are shielded with metallayers to avoid interference with the amplifiers of other stages.

A four-channel nanopore detector front-end is presented by Parsnejad et al. (2016). As shown in Figure 10, half of the differential amplifier, which receives the constant input, is shared among the different cells to reduce the consumed area and power. No ADCs are used in this sensor.

In Table 2, a comparison is presented among state-of-the-art electrochemical CMOS arrays interfaced to electrical current sensors. As can be seen therein, a number of the reported systems, such as Jafari et al. (2014) and White et al. (2018), can cover signal bandwidths (BWs) over 1 kHz,a figure close to the translocation rates being handled by existing commercial-grade nanopore (MinION row in Table 2) electrochemical cells (Bowden et al. 2019).

To-date, biopore-based constructions of commercial electrochemical cells have met with the most success for practical DNA sequencing. These sensors suffer from high capacitance effects and low output current levels although microfluidic chambers closely interfaced to CMOS have been increasingly successful at mitigating this (Magierowski et al. 2016). Owing to these challenges, usually front-ends with low-noise amplifiers and filters similar to the arrangement presented in Figure 7 will be suitable in the signal detection chain. Unfortunately, these front-ends are relatively area and power hungry and consequently present a large problem for large array implementations.

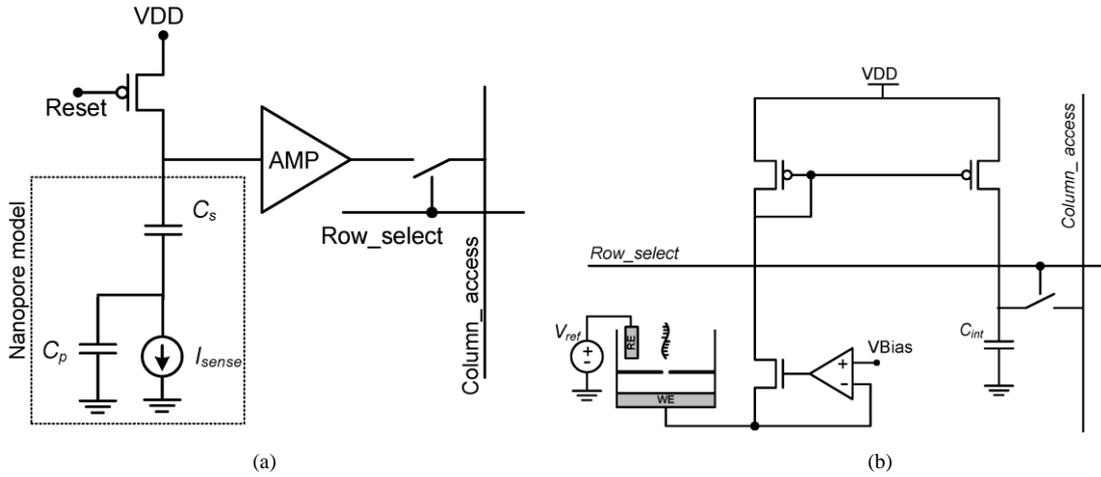

Figure 8: Examples of compact pixel designs for RPS-arrays. (a) A compact open-loop structure used for the detection of nanopore signals (Hovis et al. 2016). (b) Current mirror technique used as a compact circuit to direct the nanopore signal to the periphery circuitry (Tian 2015).

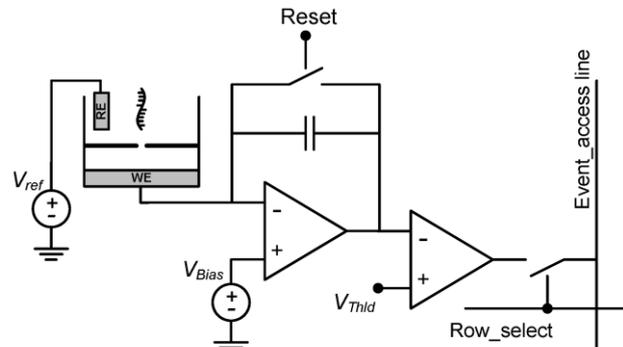

Figure 9: An event generating circuit used for nanopore signal conversion. The time-stamp at which the event occurs represents the digital pixel data (Chen 2013).

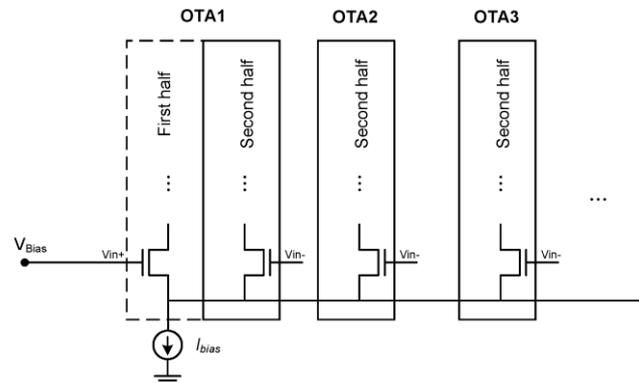

Figure 10: Half OTA sharing technique in nanopore detectors for reducing the pixel size and power (Parsnejad et al. 2016).

Table 2: A Summary of CMOS Array Approaches and Performance Levels.

| Ref., Year | Tech. [nm] | Pixel Count | Chip Area [mm$^2$] | Pixel Area [$\mu$m$^2$] | BW [Hz] | MDS [pA] | Power [mW] | AFE Type |
|---|---|---|---|---|---|---|---|---|
| Schienle et al. (2004) | 500 | 128 | 6.4×4.5 | 300$^2$ | 1 | 1 | – | In-pixel ΣΔ ADC current detector (curr. det.) |
| Gore et al. (2006) | 500 | 42 | 3×3 | 150×500 | 200 | 0.05 | 0.46 | In-pixel ΣΔ curr. det. |
| Anderson et al. (2008) | 180 | 24 | 2×4 | 300×600 | 2k | 8.5 | – | External integrating curr. det. |
| Levine et al. (2008) | 250 | 16 | 5×3 | 650$^2$ | 3k | 240 | – | In-pixel dual-slope ADC curr. det. |
| Jafari et al. (2014) | 130 | 54 | 3×3 | 200×300 | 10k | 8.6 | 0.35 | In-pixel chopper dual-slope ADC curr. det. |
| White et al. (2018) | 350 | 1024 | 5×5 | 30$^2$ | 4.4k | 0.5 | 12.5 | In-pixel curr. Integrator |
| Dong et al. (2020) | 180 | 128 | 1.2×1.3 | 1280 | 138k | 22 | 4.7 | 8 DT readout channels |
| MinION (2014) | – | 512 | 10×13 | 500$^2$ | ~1k | ~0.1 | ~1000 | In-pixel curr. Integrator |

As research devoted to the development of new pore structures advances, the electrochemical cells constructed with them benefit from lower capacitance effects and higher output currents. Since solid state nanopores promise better output signal characteristics, many workers anticipate that analog front-ends similar to those described above and also recorded in Table 2 may be used in future nanopore-based DNA sequencing systems as well.

## 6 Summary and Discussion

While the detection of specific target pathogens and DNA fragments via hybridization has been made possible using conventional electrochemical or resonance frequency measurement circuits (Rafique et al. 2019, Singh et al. 2018, Vivek et al. 2019), however design of electronics suitable for acquiring the signal of nanopore ion-channel electrochemical cells remains currently challenging due to the high frequency components of the current waveform. This paper has reviewed the status of key technologies relevant to nanopore-based molecular measurement with an emphasis on an application to DNA sequencing. Within this context, the sensor side of the technology today clearly favours RPS-based biopore sensory implementations. The determination of individual nucleotides on a DNA chain requires a high signal to noise ratio and currently PRS-based biopores have the advantage of slowing down the DNA translocation and thus reducing the signal bandwidth using biological motor molecules which attach to the DNA. A trend with this approach is to increase the throughput by using large number of channels through integrated arrays.

Although arrays of electrochemical cells integrated with CMOS readout channels have been previously presented for NAT or SBS based DNA sequencing solutions, even with the biopores, these conventional solutions are not suitable for nanopore sequencing due to the required bandwidth. Since power hungry, high bandwidth, low noise frontends are required for such DNA sequencing applications thus arrays of these frontends need to address complexities such as crosstalks, power dissipation and associated thermal issues. Another trend towards high throughput sequencing is the use of solid state nanopores equipped with local FET devices to sense traverse currents and fields. These cells promise high signal to noise ratios at high translocation rates. Research and study still continues towards engineering the suitable FET configuration which can be effective for this purpose.

## References

Anderson, E. P., Daniels, J. S., Yu, H., Karhanek, M., Lee, T. H., Davis, R. W. & Pourmand, N. (2008), 'A system for multiplexed direct electrical detection of DNA synthesis', *Sens Actuators B Chem* **129**(1), 79–86. PMC2344141[pmcid].

Ayub, M. & Bayley, H. (2016), 'Engineered transmembrane pores', *Current Opinion in Chemical Biology* **34**, 117–126. S1367-5931(16)30102-8[PII].

Bausells, J., Carrabina, J., Errachid, A. & Merlos, A. (1999), 'Ion-sensitive field-effect transistors fabricated in a commercial CMOS technology', *Sens. Actuat. B* **57**, 56–62.

Bergveld, P. (1970), 'Development of an ion-sensitive solid-state device for neurophysiological measurements', *IEEE Trans. Biomed. Eng.* **17**, 70–71.

Bowden, R., Davies, R. W., Heger, A., Pagnamenta, A. T., de Cesare, M., Oikkonen, L. E., Parkes, D., Freeman, C., Dhalla, F., Patel, S. Y., Popitsch, N., Ip, C. L. C., Roberts, H. E., Salatino, S., Lockstone, H., Lunter, G., Taylor, J. C., Buck, D., Simpson, M. A. & Donnelly, P. (2019), 'Sequencing of human genomes with nanopore technology', *Nature communications* **10**(1), 1869–1869. PMC6478738[pmcid].

Brown, C. G. & Clarke, J. (2016), 'Nanopore development at Oxford Nanopore', *Nature Biotechnology* **34**(8), 810–811.

Cadinu, P., Campolo, G., Pud, S., Yang, W., Edel, J. B., Dekker, C. & Ivanov, A. P. (2018), 'Double barrel nanopores as a new tool for controlling single-molecule transport', *Nano Letters* **18**(4), 2738–2745. PMID: 29569930.

Chen, C., Li, Y., Kerman, S., Neutens, P., Willems, K., Cornelissen, S., Lagae, L., Stakenborg, T. & Van Dorpe, P. (2018), 'High spatial resolution nanoslit sers for single-molecule nucleobase sensing', Nature communications 9(1), 1–9.

Chen, R. (2013), 'Noise shielding techniques for ultra low current measurements in biochemical applications'. US Patent 8,541,849.

Chien, C.-C., Shekar, S., Niedzwiecki, D. J., Shepard, K. L. & Drndić, M. (2019), 'Single-stranded DNA translocation recordings through solid-state nanopores on glass chips at 10 MHz measurement bandwidth', *ACS nano* **13**(9), 10545–10554.

Church, G., Deamer, D. W., Branton, D., Baldarelli, R. & Kasianowicz, J. (1998), 'Characterization of individual polymer molecules based on monomer-interface interactions'. Filed: Mar. 17, 1995.

Clarke, J., Wu, H.-C., Jayasinghe, L., Patel, A., Reid, S. & Bayley, H. (2009), 'Continuous base identification for single-molecule nanopore DNA sequencing', *Nat. Nanotech.* **4**, 265–270.

Crescentini, M., Tartagni, M., Morgan, H. & Traverso, P. A. (2017), 'A compact low-noise broadband digital picoammeter architecture', *Measurement* **100**, 194–204.

Cui, Y., Wei, Q., Park, H. & Lieber, C. M. (2001), 'Nanowire nanosensors for highly sensitive and selective detection of biological and chemical species', *Science* **293**(5533), 1289–1292.

Dai, S., Perera, R. T., Yang, Z. & Rosenstein, J. K. (2016), 'A 155-db dynamic range current measurement front end for electrochemical biosensing', *IEEE Transactions on Biomedical Circuits and Systems* **10**(5), 935–944.

Dawji, Y., Zetterblom, F., Benages, S., Morral, B., Tan, S., Sjoland, H. & Magierowski, S. (2019), A 65-nm cmos low-power front-end for 3rd generation DNA sequencing, *in* '2019 IEEE Sensors', IEEE, pp. 1–4.

Derrington, I. M., Butler, T. Z., Collins, M. D., Manrao, E., Pavlenok, M., Niederweis, M. & Gundlach, J. H. (2010), 'Nanopore DNA sequencing with MspA', Proc. Natl. Acad. Sci. 107(37), 16060–16065.

Dong, C., Jiang, Y., Jiang, K., Huang, Y. & Qin, Y. (2020), A 37.37 $\mu$W-per-cell multifunctional automated nanopore sequencing cmos platform with 16×8 biosensor array, *in* '2020 IEEE International Symposium on Circuits and Systems (ISCAS)', IEEE, pp. 1–4.

Farshad, M. & Rasaiah, J. C. (2020), 'Molecular dynamics simulation study of transverse and longitudinal ionic currents in solid-state nanopore DNA sequencing', *ACS Applied Nano Materials* **3**(2), 1438–1447.

Feng, J., Liu, K., Bulushev, R. D., Khlybov, S., Dumcenco, D., Kis, A. & Radenovic, A. (2015), 'Identification of single nucleotides in $MoS_2$ nanopores', *Nature Nanotechnology* **10**, 1070–1076.

Ferrari, G., Farina, M., Guagliardo, F., Carminati, M. & Sampietro, M. (2009), 'Ultra-low-noise cmos current preamplifier from dc to 1 MHz', *Electronics Letters* **45**(25), 1278–1280.

Ferrari, G., Gozzini, F., Molari, A. & Sampietro, M. (2009), 'Transimpedance amplifier for high sensitivity current measurements on nanodevices', *IEEE J. Solid-State Circuits* **44**(5), 1609–1616.

Ferrari, G., Gozzini, F. & Sampietro, M. (2007), A current-sensitive front-end amplifier for nano-biosensors with a 2MHz BW, *in* '2007 IEEE International Solid-State Circuits Conference. Digest of Technical Papers', pp. 164–165.

Fish, D. A. & White, S. P. (2018), 'Apparatus and methods for measuring an electrical current'. US Patent App. 15/573,332.

Fodor, S. P. A., Rava, R. P., Huang, X. C., Pease, A. C., Holmes, C. P. & Adams, C. L. (1993), 'Multiplexed biochemical assays with biological chips', *Nature* **364**(6437), 555–556.

Fologea, D., Uplinger, J., Thomas, B., McNabb, D. S. & Li, J. (2005), 'Slowing DNA translocation in a solid-state nanopore', *Nano Letters* **5**(9), 1734–1737. PMID: 16159215.

Fragasso, A., Schmid, S. & Dekker, C. (2020), 'Comparing current noise in biological and solid-state nanopores', *ACS nano* **14**(2), 1338–1349.

Fuller, C. W., Kumar, S., Porel, M., Chien, M., Bibillo, A., Stranges, P. B., Dorwart, M., Tao, C., Li, Z., Guo, W. et al. (2016), 'Real-time single-molecule electronic DNA sequencing by synthesis using polymer-tagged nucleotides on a nanopore array', *Proc. Natl. Acad. Sci.* **113**(19), 5233–5238.

Garaj, S., Hubbard, W., Reina, A., Kong, J., Branton, D. & Golovchenko, J. A. (2010), 'Graphene as a subnanometre trans-electrode membrane', *Nature* **467**, 190–193.

Gilboa, T., Zrehen, A., Girsault, A. & Meller, A. (2018), 'Optically-monitored nanopore fabrication using a focused laser beam', *Scientific Reports* **8**(1), 9765.

Goldstein, B., Kim, D., Xu, J., Vanderlick, T. K. & Culurciello, E. (2012), 'Cmos low current measurement system for biomedical applications', *IEEE Transactions on Biomedical Circuits and Systems* **6**(2), 111–119.

Gore, A., Chakrabartty, S., Pal, S. & Alocilja, E. C. (2006), 'A multichannel femtoampere- sensitivity potentiostat array for biosensing applications', *IEEE Transactions on Circuits and Systems I: Regular Papers* **53**(11), 2357–2363.

Goyal, P., Krasteva, P. V., Gerven, N. V., Gubellini, F. et al. (2014), 'Structural and mechanistic insights into the bacterial amyloid secretion channel CsgG', *Nature* **516**(7530), 250–253.

Graf, M., Liu, K., Sarathy, A., Leburton, J.-P. & Radenovic, A. (2018), 'Transverse detection of DNA in a $MoS_2$ nanopore', *Biophysical Journal* **114**(3), 180a.

Hartel, A. J., Ong, P., Schroeder, I., Giese, M. H., Shekar, S., Clarke, O. B., Zalk, R., Marks, A. R., Hendrickson, W. A. & Shepard, K. L. (2018), 'Single-channel recordings of RyR1 at microsecond resolution in CMOS-suspended membranes', *Proceedings of the National Academy of Sciences* **115**(8), E1789–E1798.

Hartel, A. J., Shekar, S., Ong, P., Schroeder, I., Thiel, G. & Shepard, K. L. (2019), 'High bandwidth approaches in nanopore and ion channel recordings-a tutorial review', *Analytica chimica acta* **1061**, 13–27.

Hassibi, A., Manickam, A., Singh, R., Bolouki, S., Sinha, R., Jirage, K. B., McDermott, M. W., Hassibi, B., Vikalo, H., Mazarei, G., Pei, L., Bousse, L., Miller, M., Heshami, M., Savage, M. P., Taylor, M. T., Gamini, N., Wood, N., Mantina, P., Grogan, P., Kuimelis, P., Savalia, P., Conradson, S., Li, Y., Meyer, R. B., Ku, E., Ebert, J., Pinsky, B. A., Dolganov, G., Van, T., Johnson, K. A., Naraghi-Arani, P., Kuimelis, R. G. & Schoolnik, G. (2018), 'Multiplexed identification, quantification and genotyping of infectious agents using a semiconductor biochip', *Nature Biotechnology* **36**(8), 738–745.

Heerema, S. J. & Dekker, C. (2016), 'Graphene nanodevices for DNA sequencing', Nat. Nanotech. 11, 127–136.

Heerema, S. J., Vicarelli, L., Pud, S., Schouten, R. N., Zandbergen, H. W. & Dekker, C. (2018), 'Probing DNA translocations with inplane current signals in a graphene nanoribbon with a nanopore', *ACS nano* **12**(3), 2623–2633.

Hovis, J., Tian, H. & Chen, R. J. (2016), 'Adjustable bilayer capacitance structure for biomedical devices'. US Patent App. 14/606,632.

Howorka, S. (2017), 'Building membrane nanopores', *Nature Nanotechnology* **12**(6), 619–630. Review Article.

Howorka, S., Remaut, H., Jayasinghe, L., Wallace, E. J., Clarke, J. A., Hambley, R. G. & Pugh, J. B. (2017), 'Mutant CsgG Pores'.

Hsu, C. & Hall, D. A. (2018), A current-measurement front-end with 160db dynamic range and 7ppm inl, *in* '2018 IEEE International Solid - State Circuits Conference - (ISSCC)', pp. 326–328.

Hsu, C., Jiang, H., Venkatesh, A. G. & Hall, D. A. (2015), 'A hybrid semi-digital transimpedance amplifier with noise cancellation technique for nanopore-based DNA sequencing', *IEEE Transactions on Biomedical Circuits and Systems* **9**(5), 652–661.

Huang, S., He, J., Chang, S., Zhang, P., Liang, F., Li, S., Tuchband, M., Fuhrmann, A., Ros, R. & Lindsay, S. (2010), 'Identifying single bases in a DNA oligomer with electron tunnelling', *Nature Nanotechnology* **5**, 868–873.

Huang, S., Romero-Ruiz, M., Castell, O. K., Bayley, H. & Wallace, M. I. (2015), 'High-throughput optical sensing of nucleic acids in a nanopore array', *Nature Nanotechnology* **10**(11), 986–991. Article.

Huang, Y., Magierowski, S. & Ghafar-Zadeh, E. (2016), Cmos for high-speed nanopore DNA basecalling, *in* 'Proc. IEEE ISCAS', pp. 1–4.

Ivanov, A. P., Instuli, E., McGilvery, C. M., Baldwin, G., McComb, D. W., Albrecht, T. & Edel, J. B. (2011), 'DNA tunneling detector embedded in a nanopore', Nano Letters 11(1), 279–285. PMID: 21133389.

Jafari, H. M., Abdelhalim, K., Soleymani, L., Sargent, E. H., Kelley, S. O. & Genov, R. (2014), 'Nanostructured CMOS wireless ultra-wideband label-free PCR-free DNA analysis SOC', *IEEE Journal of Solid-State Circuits* **49**(5), 1223–1241.

Jain, M., Koren, S., Miga, K. H., Quick, J., Rand, A. C., Sasani, T. A., Tyson, J. R., Beggs, A. D., Dilthey, A. T., Fiddes, I. T., Malla, S., Marriott, H., Nieto, T., O'Grady, J., Olsen, H. E., Pedersen, B. S., Rhie, A., Richardson, H., Quinlan, A. R., Snutch, T. P., Tee, L., Paten, B., Phillippy, A. M., Simpson, J. T., Loman, N. J. & Loose, M. (2018), 'Nanopore sequencing and assembly of a human genome with ultra-long reads', Nature Biotechnology 36(4), 338–345.

Jayasinghe, L., Wallace, E. J. & Singh, P. (2018), 'Modified Nanopores, Compositions Comprising the Same, and Uses Thereof'.

Jayasinghe, L., Wallace, E. J., Singh, P., Hambley, R. G., Jordan, M. & Remaut, H. (2018), 'Transmembrane Pore Consisting of Two CsgG Pores'.

Jiang, Y., Liu, X., Dang, T. C., Huang, X., Feng, H., Zhang, Q. & Yu, H. (2018), 'A high-sensitivity potentiometric 65-nm cmos isfet sensor for rapid e. coli screening', *IEEE Transactions on Biomedical Circuits and Systems* **12**(2), 402–415.

Kasianowicz, J. J., Brandin, E., Branton, D. & Deamer, D. W. (1996), 'Characterization of individual polynucleotide molecules using a membrane channel', *Proceedings of the National Academy of Sciences* **93**(24), 13770–13773.

Kim, D., Goldstein, B., Tang, W., Sigworth, F. J. & Culurciello, E. (2013), 'Noise analysis and performance comparison of low current measurement systems for biomedical applications', *IEEE Transactions on Biomedical Circuits and Systems* **7**(1), 52–62.

Kim, J., Maitra, R., Pedrotti, K. D. & Dunbar, W. B. (2013), 'A patch-clamp ASIC for nanopore-based DNA analysis', *IEEE Trans. Biomed. Circuits Syst.* **7**(3), 285–295.

Kowalczyk, S. W., Grosberg, A. Y., Rabin, Y. & Dekker, C. (2011), 'Modeling the conductance and DNA blockade of solid-state nanopores', *Nanotechnology* **22**, 1–5.

Kowalczyk, S. W., Grosberg, A. Y., Rabin, Y. & Dekker, C. (2012), 'Reply to comment on 'modeling the conductance and DNA blockade of solid-state nanopores'', *Nanotechnology* **23**, 1.

Kruppa, P., Frey, A., Kuehne, I., Schienle, M., Persike, N., Kratzmueller, T., Hartwich, G. & Schmitt-Landsiedel, D. (2010), 'A digital cmos-based 24× 16 sensor array platform for fully automatic electrochemical DNA detection', *Biosensors and Bioelectronics* **26**(4), 1414–1419.

Kwok, H., Briggs, K. & Tabard-Cossa, V. (2014), 'Nanopore fabrication by controlled dielectric breakdown', *PLoS One* **9**(3), e92880–e92880. PONE-D-13-51563[PII].

Lee, K., Park, K.-B., Kim, H.-J., Yu, J.-S., Chae, H., Kim, H.-M. & Kim, K.-B. (2018), 'Recent progress in solid-state nanopores', Advanced Materials 30(42), 1704680.

Levine, P. M., Gong, P., Levicky, R. & Shepard, K. L. (2008), 'Active CMOS sensor array for electrochemical biomolecular detection', IEEE Journal of Solid-State Circuits 43(8), 1859–1871.

Li, P., Molderez, T. R., Ceyssens, F., Rabaey, K. & Verhelst, M. (2019), A 64-channel, 1.1-pA-

accurate on-chip potentiostat for parallel electrochemical monitoring, *in* 'ESSCIRC 2019-IEEE 45th European Solid State Circuits Conference (ESSCIRC)', IEEE, pp. 317–320.

Liu, X., Zhang, Y., Nagel, R., Reisner, W. & Dunbar, W. B. (2018), 'Controlling DNA Tug-of-War in a Dual Nanopore Device', *arXiv e-prints* p. arXiv:1811.11105.

Long, Z., Zhan, S., Gao, P., Wang, Y., Lou, X. & Xia, F. (2018), 'Recent advances in solid nanopore/channel analysis', *Analytical Chemistry* **90**(1), 577–588. PMID: 29161021.

Lu, J. & Holleman, J. (2012), A wideband ultra-low-current on-chip ammeter, *in* 'Proceedings of the IEEE 2012 Custom Integrated Circuits Conference', pp. 1–4.

Magierowski, S., Huang, Y., Wang, C. & Ghafar-Zadeh, E. (2016), 'Nanopore-CMOS interfaces for DNA sequencing', *Biosensors* **6**(3).

Mardis, E. R. (2017), 'DNA sequencing technologies: 2006-2016', *Nature Protocols* **12**(2), 213–218. Perspective.

McCombie, W. R., McPherson, J. D. & Mardis, E. R. (2018), 'Next-generation sequencing technologies', *Cold Spring Harbor Perspectives in Medicine*.

McNally, B., Singer, A., Yu, Z., Sun, Y., Weng, Z. & Meller, A. (2010), 'Optical recognition of converted DNA nucleotides for single-molecule DNA sequencing using nanopore arrays', *Nano Letters* **10**(6), 2237–2244. PMID: 20459065.

Merchant, C. A., Healy, K., M.Wanunu, Ray, V., Peterman, N., Bartel, J., Fischbein, M. D., Venta, K., Luo, Z., Johnson, A. T. C. & Drndic, M. (2010), 'DNA translocation through graphene nanopores', *Nano Lett.* **10**(8), 4916–4924.

Merriman, B. & Rothberg, J. M. (2012), 'Progress in Ion Torrent semiconductor chip based sequencing', *Electrophor.* **33**, 3397–3417.

Milgrew, M. J., Cumming, D. R. S. & Hammond, P. A. (2003), The fabrication of scalable multi-sensor arrays using standard CMOS technology, *in* 'Proc. IEEE Custom Integrated Circuits Conf.', pp. 513–516.

Milgrew, M. J., Hammond, P. A. & Cumming, D. R. S. (2004), 'The development of scalable sensor arrays using standard CMOS technology', *Sens. Actuat. B* **103**, 37–42.

Molderez, T. R., Rabaey, K. & Verhelst, M. (2021), 'A scalable 128-channel, time-multiplexed potentiostat for parallel electrochemical experiments', *IEEE Transactions on Circuits and Systems I: Regular Papers* **68**(3), 1068–1079.

Nelson, T., Zhang, B. & Prezhdo, O. V. (2010), 'Detection of nucleic acids with graphene nanopores: Ab initio characterization of a novel sequencing device', *Nano Letters* **10**(9), 3237–3242. PMID: 20722409.

Parkin, W. M. & Drndić, M. (2018), 'Signal and noise in FET-nanopore devices', ACS Sensors 3(2), 313–319. PMID: 29322780.

Parsnejad, S., Li, H. & Mason, A. J. (2016), Compact CMOS amperometric readout for nanopore arrays in high throughput Lab-on-CMOS, *in* 'Circuits and Systems (ISCAS), 2016 IEEE International Symposium on', IEEE, pp. 2851–2854.

Prasongkit, J., Grigoriev, A., Pathak, B., Ahuja, R. & Scheicher, R. H. (2011), 'Transverse conductance of DNA nucleotides in a graphene nanogap from first principles', *Nano Letters* **11**(5), 1941–1945.

Pud, S., Chao, S.-H., Belkin, M., Verschueren, D., Huijben, T., van Engelenburg, C., Dekker, C. & Aksimentiev, A. (2016), 'Mechanical trapping of DNA in a double-nanopore system', *Nano Lett* **16**(12), 8021–8028. PMC5523128[pmcid].

Rafique, B., Iqbal, M., Mehmood, T. & Shaheen, M. A. (2019), 'Electrochemical DNA biosensors: A review', *Sensor Review*.

Ren, R., Wang, X., Cai, S., Zhang, Y., Korchev, Y., Ivanov, A. P. & Edel, J. B. (2020), 'Selective sensing of proteins using aptamer functionalized nanopore extended field-effect transistors', *Small Methods* **4**(11), 2000356.

Rodriguez-Manzo, J. A., Puster, M., Nicolai, A., Meunier, V. & Drndic, M. (2015), 'DNA translocation in nanometer thick silicon nanopores', *ACS Nano* **9**(6), 6555–6564.

Rosenstein, J. K., Ramakrishnan, S., Roseman, J. & Shepard, K. L. (2013), 'Single ion channel recordings with CMOS-anchored lipid membranes', *Nano Lett.* **13**(6), 2682–2686.

Rosenstein, J. K., Wanunu, M., Merchant, C. A., Drndic, M. & Shepard, K. L. (2012), 'Integrated nanopore sensing platform with sub-microsecond temporal resolution', *Nat. Methods* **9**(5), 487–492.

Rothberg, J. M., Hinz, W., Johnson, K. L. & Bustillo, J. (2011), 'Methods and apparatus for measuring analytes using large scale FET arrays'. Filed: Dec. 14, 2007.

Rothberg, J. M. et al. (2011), 'An integrated semiconductor device enabling non-optical genome sequencing', *Nature* **475**(7356), 348–352.

Rothe, J., Lewandowska, M., Heer, F., Frey, O. & Hierlemann, A. (2011), 'Multi-target electrochemical biosensing enabled by integrated CMOS electronics', *Journal of Micromechanics and Microengineering* **21**(5), 054010.

Schienle, M., Paulus, C., Frey, A., Hofmann, F., Holzapfl, B., Schindler-Bauer, P. & Thewes, R. (2004), 'A fully electronic DNA sensor with 128 positions and in-pixel A/D conversion', *IEEE Journal of Solid-State Circuits* **39**(12), 2438–2445.

Schneider, G. F., Kowalczyk, S. W., Calado, V. E., Pandraud, G., Zandbergen, H. W., Vandersypen, L. M. K. & Dekker, C. (2010), 'DNA translocation through graphene nanopores', Nano Letters 10(8), 3163–3167. PMID: 20608744.

Shekar, S., Chien, C.-C., Hartel, A., Ong, P., Clarke, O. B., Marks, A., Drndic, M. & Shepard, K. L. (2019), 'Wavelet denoising of high-bandwidth nanopore and ion-channel signals', Nano letters 19(2), 1090–1097.

Shekar, S., Niedzwiecki, D. J., Chien, C.-C., Ong, P., Fleischer, D. A., Lin, J., Rosenstein, J. K., Drndi´c, M. & Shepard, K. L. (2016), 'Measurement of DNA translocation dynamics in a solid-state nanopore at 100 ns temporal resolution', *Nano Letters* **16**(7), 4483–4489. PMID: 27332998.

Shendure, J., Balasubramanian, S., Church, G. M., Gilbert, W., Rogers, J., Schloss, J. A. & Waterston, R. H. (2017), 'DNA sequencing at 40: past, present and future', *Nature* **550**(7676), 345–353. Review Article.

Shi, W., Friedman, A. K. & Baker, L. A. (2017), 'Nanopore sensing', *Analytical Chemistry* **89**(1), 157–188. PMID: 28105845.

Shi, X., Verschueren, D. V. & Dekker, C. (2018), 'Active delivery of single DNA molecules into a plasmonic nanopore for label-free optical sensing', *Nano Letters* **18**(12), 8003–8010.

Sigworth, F. J. (1995), Electronic design of the patch clamp, *in* B. Sakmann & E. Neher, eds, 'Single-Channel Recording', 2 edn, Springer, New York.

Singh, S., Kaushal, A. & Kumar, A. (2018), 'Recent advances in sensors for early detection of pathogens causing rheumatic heart disease', *Sensor Review*.

Song, L. Z., Hobaugh, M. R., Shustack, C., Cheley, S., Bayley, H. & Gouaux, J. E. (1996), 'Structure of staphylococcal alpha-hemolysin, a heptameric transmembrane pore', *Science* **274**, 1859–1866.

Soni, G. V. & Meller, A. (2007), 'Progress toward ultrafast DNA sequencing using solid-state nanopores', *Clinical Chemistry* **53**(11), 1996–2001.

Stoddart, D., Maglia, G., Mikhailova, E., Heron, A. J. & Bayley, H. (2010), 'Multiple base-recognition sites in a biological nanopore: two heads are better than one', *Angew Chem Int Ed Engl* **49**(3), 556–559. PMC3128935[pmcid].

Sutula, S., Pallarés Cuxart, J., Gonzalo-Ruiz, J., Muñoz-Pascual, F. X., Terés, L. & Serra-Graells, F. (2014), 'A 25-μW all-MOS potentiostatic delta-sigma ADC for smart electrochemical sensors', IEEE Transactions on Circuits and Systems I: Regular Papers 61(3), 671–679.

Taherzadeh-Sani, M., Hussaini, S. M. H., Rezaee-Dehsorkh, H., Nabki, F. & Sawan, M. (2017), 'A 170-db ω cmos tia with 52-pa input-referred noise and 1-mhz bandwidth for very low current sensing', IEEE Transactions on Very Large Scale Integration (VLSI) Systems 25(5), 1756–1766.

Tedjo, W. & Chen, T. (2019), 'An integrated biosensor system with a high-density microelectrode array for real-time electrochemical imaging', *IEEE transactions on biomedical circuits and systems* **14**(1), 20–35.

Tedjo, W., Nejad, J. E., Feeny, R., Yang, L., Henry, C. S., Tobet, S. & Chen, T. (2018), 'Electrochemical biosensor system using a CMOS microelectrode array provides high spatially and temporally resolved images', *Biosensors and Bioelectronics* **114**, 78–88.

Thewes, R., Hofmann, F., Frey, A., Holzapfl, B., Schienle, M., Paulus, C., Schindler, P., Eckstein, G., Kassel, C., Stanzel, M., Hintsche, R., Nebling, E., Albers, J., Hassman, J., Schulein, J., Goemann, W. & Gumbrecht, W. (2002), Sensor arrays for fully-electronic DNA detection on CMOS, *in* 'ISSCC Dig. Tech. Papers', Vol. 1, pp. 350–473 vol.1.

Tian, H. (2015), 'Nanopore-based sequencing chips using stacked wafer technology'. US Patent App. 14/225,263.

Traversi, F., Raillon, C., Benameur, S. M., Liu, K., Khlybov, S., Tosun, M., Krasnozhon, D., Kis, A. & Radenovic, A. (2013), 'Detecting the translocation of DNA through a nanopore using graphene nanoribbons', Nature Nanotechnology 8(12), 939–945. Article.

Tsutsui, M., Taniguchi, M., Yokota, K. & Kawai, T. (2010), 'Identifying single nucleotides by tunnelling current', *Nature Nanotechnology* **5**, 286–290.

Venkatesan, B. M. & Bashir, R. (2011), 'Nanopore sensors for nucleic acid analysis', *Nature Nanotechnology* **6**(10), 615–624.

Venta, K., Shemer, G., Puster, M., Rodriguez-Manzo, J. A., Balan, A., Rosenstein, J. K., Shepard, K. & Drndic, M. (2013), 'Differentiation of short, single-stranded DNA homopolymers in solid- state nanopores', ACS Nano 7(5), 4629–4636.

Ventra, M. D. & Taniguchi, M. (2016), 'Decoding DNA, RNA and peptides with quantum tunnelling', *Nat. Nanotech.* **11**, 117–126.

Verschueren, D. V., Pud, S., Shi, X., De Angelis, L., Kuipers, L. & Dekker, C. (2018), 'Label-free optical detection of DNA translocations through plasmonic nanopores', *ACS Nano* pp. 1–10.

Vivek, A., Shambavi, K. & Alex, Z. C. (2019), 'A review: Metamaterial sensors for material characterization', *Sensor Review*.

Wen, C. & Zhang, S.-L. (2020), 'Fundamentals and potentials of solid-state nanopores: a review', Journal of Physics D: Applied Physics 54(2), 023001.

White, K. A., Mulberry, G. & Kim, B. N. (2018), 'Rapid 1024-pixel electrochemical imaging at 10,000 frames per second using monolithic CMOS sensor and multifunctional data acquisition system', *IEEE Sensors Journal* **18**(13), 5507–5514.

Willmott, G. R. & Smith, B. G. (2012), 'Comment on 'modeling the conductance and DNA blockade of solid-state nanopores'', *Nanotechnology* **23**, 1.

Xie, P., Xiong, Q., Fang, Y., Qing, Q. & Lieber, C. M. (2012), 'Local electrical potential detection of DNA by nanowire-nanopore sensors', *Nature Nanotechnology* **7**(2), 119–125.

Yamazaki, H., Hu, R., Zhao, Q. & Wanunu, M. (2018), 'Photothermally assisted thinning of silicon nitride membranes for ultrathin asymmetric nanopores', *ACS Nano* **12**(12), 12472–12481. PMID: 30457833.

Yang, C., Huang, Y., Hassler, B. L., Worden, R. M. & Mason, A. J. (2009), 'Amperometric electrochemical microsystem for a miniaturized protein biosensor array', *IEEE Transactions on Biomedical Circuits and Systems* **3**(3), 160–168.

Zhang, Y., Miyahara, Y., Derriche, N., Yang, W., Yazda, K., Liu, Z., Grutter, P. & Reisner, W. (2019), 'Nanopore fabrication via tip-controlled local breakdown using an atomic force microscope', *arXiv e-prints* p. arXiv:1901.07071.

Zwolak, M. & Di Ventra, M. (2005), 'Electronic signature of DNA nucleotides via transverse transport', *Nano Letters* **5**(3), 421–424. PMID: 15755087.